\documentclass[12pt]{iopart}

\usepackage{iopams}
\usepackage{graphicx}
\usepackage{eurosym}
\usepackage{setspace}
\usepackage{array}
\usepackage{hyperref}
\usepackage{xcolor}

\begin{document}

\title{Quantum-droplet interferometry}

\author{Sriganapathy Raghav}
	    \address{Department of Physics, Indian Institute of Technology Patna, Bihta, Patna-801103, India}
\author{Boris Malomed}
		\address{Department of Physical Electronics, School of Electrical Engineering, Faculty of Engineering, and The Center for Light-Matter Interaction, Tel Aviv University, Tel Aviv 69978, Israel}
\author{Utpal Roy \footnote{Corresponding author}}
		\address{Department of Physics, Indian Institute of Technology Patna, Bihta, Patna-801103, India}

\ead{uroy@iitp.ac.in}

\begin{abstract}
We propose atom interferometers based on quantum droplet (QD), which is also being reported as a superior platform for interferometry. The emphasizes have been given to harmonic-oscillator (HO) or ring-shaped potentials. In the HO trap, a
Gaussian barrier induces coherent splitting; in the ring, one or two
barriers guide the splitting and subsequent recombination. The atom number
and relative mean-field interaction strength critically affect the interferometric performance. The transmission-coefficient analysis identifies values of the barrier parameters for the balanced $50:50$ splitting. The
post-recombination atom-number imbalance serves as a sensitive indicator of
the relative phase between merging daughter QDs. We demonstrate that the
HO-based setup may serve as a tilt-meter and target detector, and the ring
geometry may be used as a compact QD Sagnac interferometer for rotation
sensing.
\end{abstract}

\maketitle

\pagenumbering{arabic}


\section{Introduction}

Quantum droplets (QDs) are self-bound states of Bose-Einstein condensates
(BECs) stabilized by the balance between the mean-field (MF) and
beyond-mean-field (BMF) interactions \cite{petrov2015quantum,luo2021new}.
The initial motivation for studying QDs was the search for stable
self-trapped BEC configurations in two and three dimensions (2D and 3D). In
the high-dimensional settings, the MF theory predicts the collapse of BEC
with attractive interactions \cite{pethick2008bose,pitaevskii2016bose,MalomedMS}. However, the BMF (Lee-Huang-Yang or LHY) \cite%
{lee1957eigenvalues}) correction to the corresponding Gross-Pitaevskii
equations (GPEs) may arrest the collapse, enabling the formation of stable
QDs. As proposed by Petrov \cite{petrov2015quantum}, the careful tuning of
inter-species and intra-species interactions in a binary BEC leads to a
modified GPE that incorporates both MF and BMF terms with opposite signs.
The BMF term (alias the LHY correction) accounts for quantum fluctuations
around the MF states, which modify the system's ground state. In the 3D
LHY-amended GPE, the cubic term, which represents the MF interactions, is
attractive, while the quartic one, representing the LHY correction, is
repulsive \cite{petrov2015quantum}. The strength of the attractive term can
be independently controlled, allowing the formation of stable QDs, which
are filled by a dilute superfluid. Experimentally, 3D QDs were first created
in single-component BECs with dipole-dipole interactions \cite%
{kadau2016observing,ferrier2016observation,chomaz2016quantum,chomaz2019long}%
, and then in binary BEC mixtures with contact interactions \cite%
{cabrera2018quantum,semeghini2018self,cheiney2018bright,d2019observation}.

The existence, stability, and dynamics of QDs in lower dimensions have also
been extensively explored \cite{petrov2016ultradilute,astrakharchik2018dynamics,otajonov2019stationary,parisi2019liquid,parisi2020quantum,mithun2020modulational,debnath2021investigation}. The dimensionality of the system significantly affects the sign and form of the BMF term in the respective GPE, as the spectrum of zero-point
excitations depends on the density of states. In particular, in 1D, the LHY
term is quadratic and attractive, suggesting to choose the two-body MF
interaction with the repulsive\ sign for the balance \cite{petrov2016ultradilute}. Physics of 1D droplets has been widely studied in various contexts, including collective excitations \cite{tylutki2020collective}, Rabi and spin-orbit couplings \cite{chiquillo2019low,singh2020modulational,xiong2021self}, trapping in external potentials \cite{englezos2023correlated,zezyulin2023quasi,du2023ground,flynn2024harmonically,pathak2022dynamics,nie2023spectra,zhou2019dynamics,zhao2021discrete}, collision and scattering dynamics \cite{astrakharchik2018dynamics,debnath2023interaction}, and hydrodynamics \cite{chandramouli2024dispersive}. With reduced three-body losses and the absence of instabilities, such as the instability against transverse snaking \cite{kevrekidis2004avoiding}, 1D droplets exhibit stronger robustness than their 3D counterparts, which makes them a promising platform for the realisation of quantum sensing and precision measurements.

On the other hand, much interest was drawn to the use of bright solitons in
BECs with attractive MF interaction for the realisation of matter-wave
interferometry. The stability of the solitons naturally enables longer
interrogation times, making them well-suited modes for the interferometric
studies \cite%
{helm2014splitting,marchant2013controlled,mcdonald2014bright,wales2020splitting}. The MF analysis of the soliton-based interferometry setups has shown their high sensitivity to small phase variations \cite{helm2012bright,polo2013soliton,cuevas2013interactions,helm2015sagnac,sakaguchi2016matter,sun2014mean}. However, the detailed analysis including quantum noise reveals that nonlinear soliton-barrier interactions enhance relative atom number fluctuations, ultimately cancelling the improved sensitivity at high nonlinearities \cite{martin2012quantum,haine2018quantum}. On the other hand, it has been shown that the interferometry with noninteracting BECs, making the split matter-wave clouds spatially distinguishable throughout the entire
interferometric cycle, may offer better sensitivity than the soliton-based scheme \cite{haine2018quantum}.

Our objective is to elaborate a scheme for the matter-wave interferometry with QDs, rather than MF solitons, considering the generic property of the QDs, where a QD composed of a large number of atoms  ($\mathcal{N}$) features the flat-top shape. To the best of our knowledge, QD-based interferometers are not previously reported and thus, we develop schemes for interferometers in which QDs move in ring-shaped or harmonic-oscillator (HO) traps. We focus on two key parameters of QDs, \textit{viz}., the atom number $\mathcal{N}$ and MF interaction strength, $\gamma$, defined relative to the LHY one, which play a crucial role in optimizing the interferometric performance. These parameters determine the size of the QDs used in our ring-shaped interferometry and govern the droplet-to-HO determined bound state crossover in the HO-confined QDs. Building on this understanding, we propose two interferometric setups, in which splitters (potential barriers) are combined with the HO or ring-shaped traps. As in any barrier-based setup, the first collisions of the wave packet (QD, in our case) with the barrier induces the $50:50$ splitting, while the second interaction facilitates the interference-mediated recombination. The final atom population on either side of the barrier after recombination is determined by the phase shift between the merging packets, which is the main principle of the interferometry. We analyze the sensitivity of this setup to the placement of the barrier, which affects the phase difference between the recombining droplets. Further, we study the effects of HO tilt and target presence on the HO interferometer, as well as the impact of rotation on the outputs of the ring interferometer. The sensitivity of the interferometer is quantified through the contrast and relative atom-number oscillations in the output, identifying parameter regimes for optimal interferometric performance. 

The following presentation is organised as follows. In Sec.~\ref{SecII}, we
introduce the QD theoretical model; namely, the LHY-amended 1D GPE for free space is presented in subsection~\ref{IIA}, while
subsection~\ref{IIB} covers the LHY-amended 1D GPE with the HO trap. The significance of Quantum droplets for Interferometry is discussed in Section \ref{IIIF}. Section \ref{SecIII} reports findings for the QD interferometry: basic results for the scheme with the HO trap are reported in subsection~\ref{IIIA}, effects
of the HO tilt are considered in subsection~\ref{IIIB}, HO-trapped QD interferometer for target detection is discussed in subsection~\ref{IIIC}, the ring-shaped interferometer is examined in subsection \ref{IIID} and, subsection~\ref{IIIE} addresses the Sagnac interferometry. The paper is concluded by Sec.~\ref{SecIV}.
\begin{figure*}[th]
	\centering
	\includegraphics[width=16 cm]{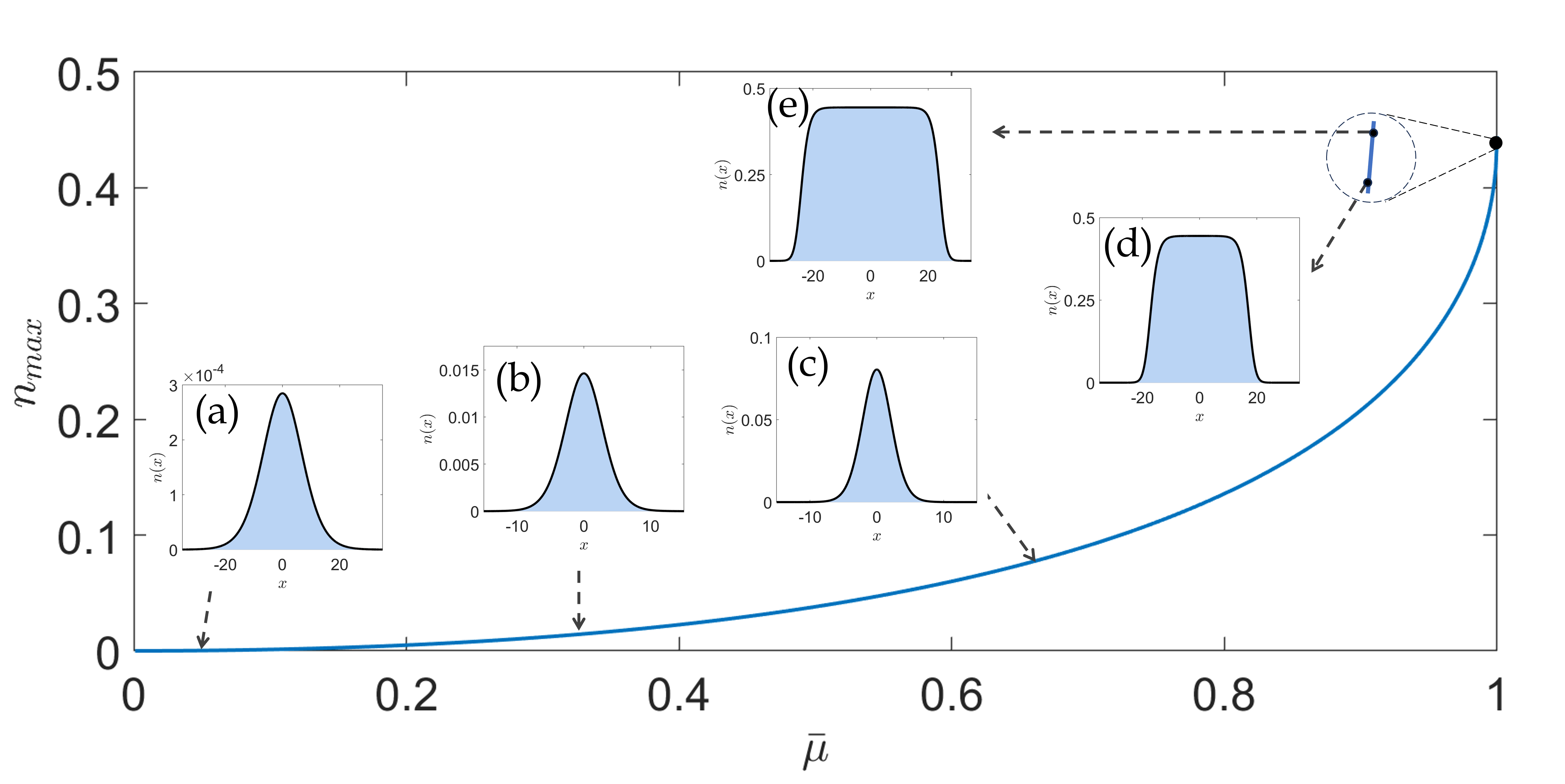}
	\caption{The maximum density of the free-space QD solution (Eq.~\ref{QDsol}), $n_{max}\equiv|\psi(x=0)|^2$, as a function of the normalized chemical potential $\bar{\protect\mu}\equiv \protect\mu /\protect\mu _{0}$ ($\protect\mu _{0}$ is the critical chemical potential) for $\protect\gamma =1$. Insets (a)-(e) exhibit the QD density profiles corresponding to $\bar{\protect\mu}=0.05$,\ $0.33$,\ $0.67$,\ $1-10^{-10}$ and $1-10^{-14}$, respectively. In this case, $x$ and $t$ are mesured in units of $l_{\perp}=0.72\;\mathrm{\protect\mu }$m and, $1/\protect\omega _{\perp }=0.32\;$ms, respectively.}
	\label{Nmax}
\end{figure*}

\section{Models for Quantum Droplets}\label{SecII}

\subsection{Quantum droplets in free space}

\label{IIA} We consider a binary BEC with equal atomic masses $m$ of the two
components and equal numbers of atoms $\ \mathcal{N}$. The 1D approximation
is maintained by the tight transverse confinement with the respective
frequency, $\omega _{\perp }$. Taking into regard the effect of quantum
fluctuations, the binary BEC is modeled by the LHY-amended system of coupled
Gross-Pitaevskii equations \cite{petrov2016ultradilute,otajonov2019stationary}:
\begin{eqnarray}
i\hbar \frac{\partial \Psi _{i}(x,t)}{\partial t} =
\Bigg[ -\frac{\hbar ^{2}}{2m}\frac{\partial ^{2}}{\partial x^{2}} 
&+& \sum_{j=1}^{2} \Gamma_{ij} |\Psi_{j}(x,t)|^{2} \nonumber\\
&& \quad - \Delta \sqrt{ \sum_{j=1}^{2} |\Psi_{j}(x,t)|^{2} } \Bigg] \Psi_{i}(x,t).\label{2GPE}
\end{eqnarray}
Here, the interaction parameters, defined as $\Gamma _{12}=(3g+g_{c})/2$, $%
\Gamma _{21}=(g_{c}-g)/2$ and $\Delta =\sqrt{2m}g^{\frac{3}{2}}/\left( \pi
\hbar \right) $, are related to intra-species repulsion coefficients, $%
g_{11}=g_{22}\equiv g>0$ and inter-species attraction ones, $%
g_{12}=g_{21}\equiv g_{c}<0$. Thus, in the symmetric case, with $\Psi
_{1}(x,t)=\Psi _{2}(x,t)\equiv \Psi (x,t)$, Eq.~(\ref{2GPE}) simplifies to
\begin{eqnarray}
    i\hbar \frac{\partial \Psi (x,t)}{\partial t}=-\frac{\hbar ^{2}}{2m}\frac{\partial ^{2}\Psi (x,t)}{\partial x^2}&+&\delta g|\Psi (x,t)|^{2}\Psi (x,t)\nonumber\\
    && \quad -\frac{\sqrt{2m}g^{\frac{3}{2}}}{\pi \hbar }|\Psi(x,t)|\Psi (x,t) \label{QDGPE},
\end{eqnarray}
where $\delta g\equiv g+g_{c}$. The wave function determines the total
number of atoms, $\mathcal{N}=\int_{-\infty }^{+\infty }\mathit{dx}|\Psi
(x,t)|^{2}$. Equation~(\ref{QDGPE}) is normalized by scaling the
coordinate, time and energy as:
\begin{equation}
     x=\bar{x}l_{\perp },\;\;t=\bar{t}/\omega _{\perp},\;\;E=\bar{E}\hbar \omega_{\perp },
\end{equation}
where $\omega _{\perp }$ and $l_{\perp }=\sqrt{\hbar /\left( m\omega _{\perp}\right)}$ are the above-mentioned transverse trapping frequency and the respective length, respectively. The scaled equation is written as
\begin{eqnarray}
	i\frac{\partial \bar{\psi}(\bar{x},\bar{t})}{\partial \bar{t}}=-\frac{1}{2}\frac{\partial ^{2}\bar{\psi}(\bar{x},\bar{t})}{\partial\bar{x}^2}&+&\gamma|\bar{\psi}(\bar{x},\bar{t})|^{2}\bar{\psi}(\bar{x},\bar{t})-|\bar{\psi}(\bar{x},\bar{t})|\bar{\psi}(\bar{x},\bar{t}),  \label{1DQDGPE}
\end{eqnarray}
where
\begin{equation}
       \gamma \equiv \frac{\pi^{2} l_{\perp}^{2}}{8a_{11}^{2}}\frac{\delta g}{g}=\frac{\pi ^{2}l_{\perp}^{2}}{8a_{11}^{2}}\left( 1+\frac{g_{c}}{g}\right) ,\;\bar{\psi}(\bar{x},\bar{t})=\frac{\sqrt{2m}g^{3/2}}{\pi \hbar ^{2}\omega
		_{\perp }}\Psi(x,t).
\end{equation}
The overbars, which are used to denote the scaled variables, are omitted
below. The cubic nonlinearity coefficient $\gamma $ quantifies the relative
MF interaction strength with respect to its BMF counterpart. As is known,
Eq.~(\ref{1DQDGPE}) give rise to the free-space ground state in the form of the QD \cite{petrov2016ultradilute},
\begin{equation}
	\psi (x,t)=\frac{-3\mu e^{-i\mu t}}{\Bigg(1+\sqrt{1+9\mu \gamma /2}\cosh
		\left( \sqrt{-2\mu }x\right) \Bigg)},  \label{QDsol}
\end{equation}%
with chemical potential $\mu $. The norm of the wave function (\ref{QDsol})
is given by
\begin{equation}
	N=n_{0}\sqrt{\frac{-2}{\mu }}\Bigg[\ln {\Bigg(\frac{1+\mu /\mu _{0}}{\sqrt{1-\mu /\mu _{0}}}\Bigg)}-\frac{\mu }{\mu _{0}}\Bigg],  \label{Number}
\end{equation}%
the corresponding number of atoms being $\mathcal{N}=\left( \pi l_{\perp
}^{3}/16a_{11}^{3}\right) N$. In Eq.~(\ref{Number}), $n_{0}=4/\left( 9\gamma
^{2}\right) $ is the equilibrium density of the spatially uniform state $%
(N\rightarrow \infty )$ and 
\begin{equation}
\mu _{0}=-2/\left( 9\gamma \right) \label{mu0}
\end{equation}
is the corresponding chemical potential. 

The current scaling is applicable in the case of $\gamma >0$, being different for $\gamma =0$ and $\gamma <0$ \cite{hu2023scattering}. As we deal with $\gamma >0$, it is relevant to examine the limit cases of $\gamma \rightarrow 0$ and $\gamma \rightarrow \infty $. In the former case, the BMF effects completely dominate over the MF ones, leading to the LHY-superfluid regime ($\gamma =0$) \cite{jorgensen2018dilute,skov2021observation}. To understand the behavior at $\gamma \rightarrow \infty $, we analyze the nonlinear interaction potential in Eq.~(\ref{1DQDGPE}):
\begin{equation}
	U_{I}(x)=\gamma n-\sqrt{n},~n\equiv |\psi (x)|^{2}. \label{NLU}
\end{equation}%
Substituting $n$ from Eq.~(\ref{QDsol}), we obtain
\begin{equation}
	U_{I}(x)=\frac{9\gamma \mu ^{2}-3|\mu |\Big(1+\sqrt{1+9\mu \gamma /2}\cosh (%
		\sqrt{-2\mu }x)\Big)}{\Big(1+\left( 9\mu \gamma /2\right) \cosh (\sqrt{-2\mu
		}x)\Big)^{2}}.  \label{U}
\end{equation}%
Setting $\mu \equiv \bar{\mu}\mu _{0}$, with $\mu _{0}$ taken as per Eq.~(\ref{mu0}) and $0<\bar{\mu}<1$, where increase of $\bar{\mu}$ implies the increase of the number of atoms, Eq. (\ref{U}) yields
\begin{equation}
	U_{I}(0)=\frac{2\bar{\mu}\Big(2\bar{\mu}-3(1+\sqrt{1-\bar{\mu}})\Big)}{%
		9\gamma (1+\sqrt{1-\bar{\mu}})^{2}}.  \label{NI}
\end{equation}
As seen from Eq. (\ref{NI}), the limit of $\gamma \rightarrow \infty$ implies that the nonlinear interaction potential vanishes at the QD center. Also from Eq.~(\ref{NLU}), in the limit of large atom number $N\rightarrow\infty$, the nonlinear interaction potential saturates to the equilibrium chemical potential $\mu_0$.
\begin{figure*}[th]
	\centering
	\includegraphics[width=16 cm]{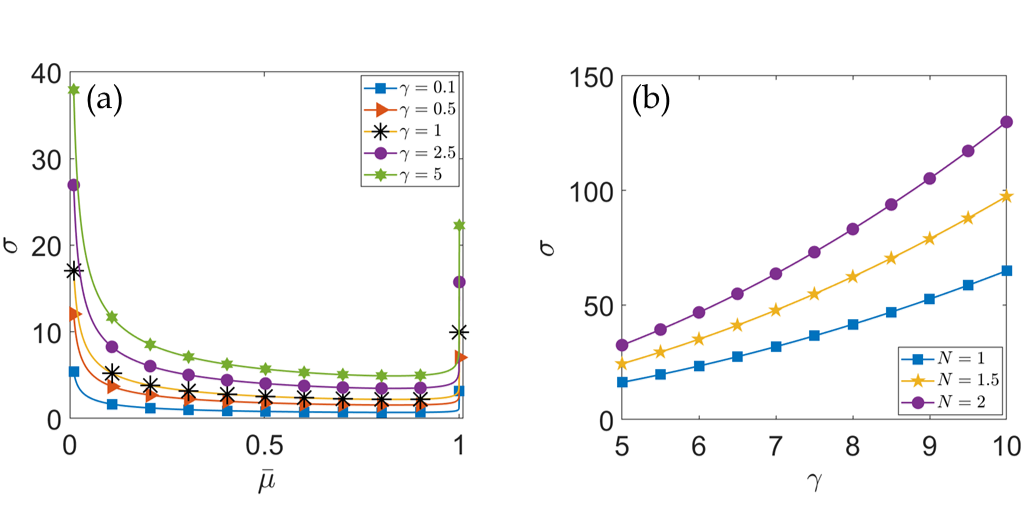}
	\caption{(a) The dependence of the RMS width, $\protect\sigma $, of the
		free-space QD solution (\protect\ref{QDsol}) on the normalized chemical potential for fixed values of the relative MF interaction strength $\protect\gamma $. (b) The dependence of $\protect\sigma $ on $\protect\gamma $ for fixed values of the atom number $N$. In this case, $x$ is measured in units of $l_{\perp}=0.72\;\mathrm{\protect\mu}$m, $t$ in units of $1/\protect\omega _{\perp }=0.32\;$ms, and $N=1$ corresponds to $1.8\times 10^{4}$ atoms.}
	\label{QD_width}
\end{figure*}

Figure \ref{Nmax} displays the dependence of the maximum QD density, $%
n_{\max }$, on the normalized chemical potential, $\bar{\mu}$, for the QD solution (Eq.~(\ref{QDsol})), with $\gamma
=1 $. At $\bar{\mu}\rightarrow 1$, $n_{\max }$ approaches the saturation
value, $n_{0}$. In this regime, the QD exhibits a flat-top profile and
becomes broader with the increase of $N$. Conversely, at $\bar{\mu}
\rightarrow 0$, $n_{\max }$ decreases towards zero, with the QD assuming a
soliton-like profile. Insets in Fig.~\ref{Nmax} display the QD density at
selected values of $\bar{\mu}$. Note that the QD width diverges in the limits
of $\bar{\mu}\rightarrow 1$ and $\bar{\mu}\rightarrow 0$, attaining a minimum at intermediate values of $\bar{\mu}$. The respective dependence of the QD's
RMS width on the chemical potential is plotted in Fig. \ref{QD_width}(a). It
is also seen that, for a given chemical potential, the width increases with
the increase of $\gamma $. In the flat-top regime, the RMS width is related
to $N$ and $\gamma $ as $\sigma \approx 9N\gamma ^{2}/\left( 8\sqrt{3}\right) $
\cite{astrakharchik2018dynamics}. Further, the dependence of $\sigma $ on $%
\gamma $ for fixed values $N=1$, $1.5$, and $2$ is plotted in Fig. \ref%
{QD_width}(b).

\begin{figure*}[th]
	\centering
	\includegraphics[width=16 cm]{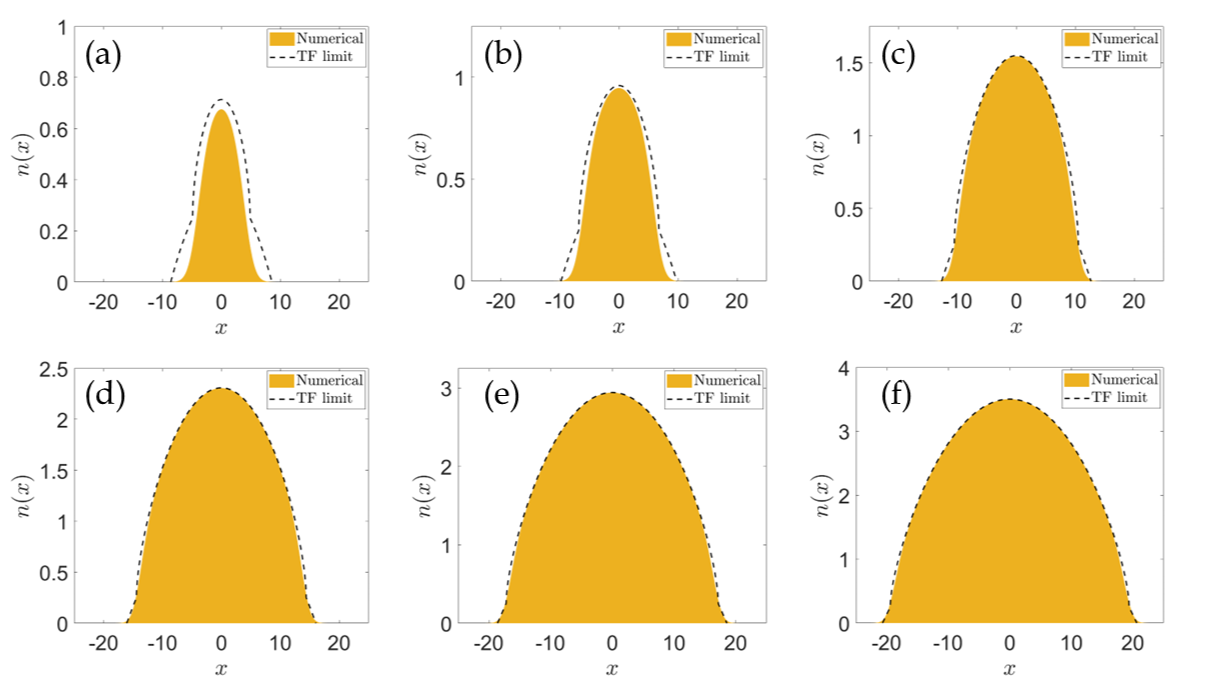}
	\caption{ The comparison of the QD density profiles for $\protect\gamma =1$, as obtained from the numerical solution and produced by the TF approximation (\protect\ref{TF}) under the action of the HO potential, for different norms: (a) $N=5$, (b) $N=10$, (c) $N=25$, (d) $N=50$, (e) $N=75$, (f) $N=100$. Here, the frequency of the HO trap is $\protect\omega_{x}=0.1\protect\omega_{\perp }$, $x$ is measured in units of $l_{\perp }=0.72\;\mathrm{\protect\mu }$m, $t$ in units of $1/\protect\omega _{\perp }=0.32\;$ms ,and $N=1$ corresponds to $1.8\times 10^{4}$ atoms.}
	\label{HQD_TF}
\end{figure*}
\begin{figure*}[th]
	\centering
	\includegraphics[width=16 cm]{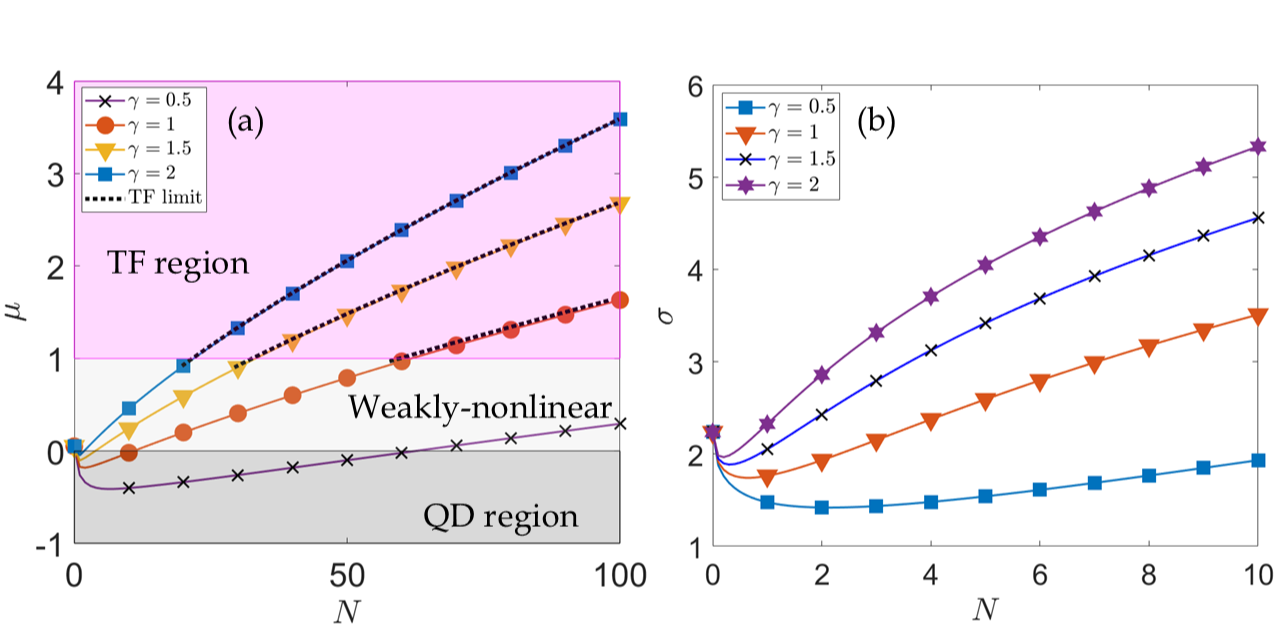}
	\caption{(a) The chemical potential $\protect\mu \ $vs. the norm $N$ for
		different values of $\protect\gamma $, in the presence of the HO potential.
		Here, we dsitinguish two regions: the QD area for $\protect\mu <0$ and the
		TF region for $\protect\mu >1$, which are separated white region of weakly
		nonlinear states. Dotted line are $\protect\mu (N)$ dependences obtained in
		the TF limit. (b) The RMS widths vs. $N$, for different values of $\protect%
		\gamma $. Here, the frequency of the HO trap is fixed as $\protect\omega %
		_{x}=0.1\protect\omega _{\perp }$. $x$ is measured in units of $l_{\perp
		}=0.72\;\mathrm{\protect\mu }$m, $t$ in units of $1/\protect\omega _{\perp
		}=0.32\;$ms, $\protect\mu $ in the unit of $\hbar \protect\omega _{\perp }$,
		and $N=1$ corresponds to $1.8\times 10^{4}$ atoms.}
	\label{mu_N}
\end{figure*}

\subsection{Harmonically trapped QDs}

\label{IIB} Here, we supplement the LHY-corrected equation~(\ref{1DQDGPE})
for QDs in free space by the HO potential:
\begin{equation}
	V(x)=\frac{1}{2}\lambda ^{2}x^{2},  \label{HP}
\end{equation}%
where, $\lambda =\omega _{x}/\omega _{\perp }$ is the
longitudinal/transverse ratio of the trapping frequencies. As usual, we assume $\lambda \ll 1$, to ensure that the setting is nearly one-dimensional
\cite{bera2022quantum,raghav2025nonlinearity}. We have obtained stable ground states of BEC trapped in the HO potential with
 $\omega _{x}=0.1\omega _{\perp }$, using the
imaginary-time-propagation method \cite%
{lehtovaara2007solution,raghav2024dispersion}. It was implemented by means
of the split-step Fourier algorithm, which handles the linear and nonlinear
parts of the GPE in the momentum and coordinate spaces, respectively. The
numerically obtained density profiles of stable QDs for $\gamma =1$ are
represented by yellow shaded regions in Fig.~\ref{HQD_TF} for different
values of the norm: (a) $N=5$, (b) $N=10$, (c) $N=25$, (d) $N=50$, (e) $N=75$
and, (f) $N=100$. The dashed lines in the same plots indicates to the
Thomas-Fermi (TF) limit, which omits the the kinetic-energy term in the GPE:
\begin{equation}
n_{\mathrm{TF}} =
\left\{
\begin{array}{ll}
(2\gamma)^{-2} \left[1 + \sqrt{1 - 2\gamma (\lambda^2 x^2 - 2\mu)} \right]^2, & |x| \leq R_{TF}, \\
0, & \mathrm{otherwise}.
\end{array}
\right.
\label{TF}
\end{equation}
The TF wavefunction, confined by the TF radius, $R_{TF}= \frac{1}{\lambda}\sqrt{2\mu-\frac{1}{2g}}$,
is plotted in Fig.~\ref{HQD_TF} for $\gamma =1$ and different values of $%
N$. The dashed lines represents the TF density profiles, which are close to
the numerical solutions for larger $N$, as might be expected. \cite%
{zezyulin2023quasi}.

In the free space, the chemical potential is related to the
number of atoms as per Eq.~(\ref{Number}), where it asymptotically
approaches $\mu _{0}$ at $N\rightarrow \infty $. The negative value of the
chemical potential signifies that the state is self-bound. The HO trap
alters this dependence, leading to
\begin{equation}
	\mu =\int_{-\infty }^{+\infty }\psi ^{\ast }\Big[-\frac{1}{2}\frac{\partial
		^{2}}{\partial x^{2}}+\gamma |\psi |^{2}-|\psi |+\frac{1}{2}\lambda ^{2}x^{2}%
	\Big]\psi \mathit{d}x .
\end{equation}%
The resultant negative and positive values of $\mu $ indicate the self-bound
QD and weakly-nonlinear (for weak nonlinearity) or quasi-TF (for strong
nonlinearity) bound states, respectively. The corresponing dependences $\mu (N)$ are plotted, for different values of $\gamma $, in Fig.~\ref{mu_N}(a). We have divided
the $\left( N,\mu \right) $ plane in two parts: $1)$ the QD regime ($\mu <0$%
, shaded green) and, $2)$ the quasi-TF regime ($\mu >1$, shaded purple) which
are separated, roughly speaking, by the white region of weakly-nonlinear
states. For fixed $\gamma $, it is observed that $\mu $ initially decreases with the growth of $N$, as long as it remains small. Beyond a certain threshold, $\mu $ starts to increase, eventually becoming positive. The transition from $\mu <0$ to $%
\mu >0$ marks the crossover from
the QDs to bound states determined by the HO trapping potential. As seen in Fig.~\ref{mu_N}(a), the crossover occurs at lower values of $N$ for larger $\gamma $, and
for $\gamma $ exceeding a certain threshold value $\gamma _{\mathrm{thr}}$,
the chemical potential remains positive for all $N$, indicating the absence
of self-bound QD states in this regime. For the HO frequency used in our
work, $\lambda =0.1$, the numerically found threshold value is $\gamma _{%
	\mathrm{thr}}\approx 4.7$. Dependences $\mu (N)$, as predicted by the TF
approximation, are shown for relevant (sufficiently large) values of $N$,
by dotted lines in Fig.~\ref{mu_N}(a), where this approximation produces
very accurate results.

The numerically identified RMS width of the HO-trapped QD is plotted, as a
function of $N$, in Fig.~\ref{mu_N}(b). For small $N$, the width decreases
with the increase of $N$, and it starts to increase above a certain
threshold value of $N$. This growth of $\sigma (N)$ becomes more pronounced
for larger values of $\gamma $. Additionally, for fixed $N$, the RMS width
increases with the growth of $\gamma $, resembling the behavior observed for the free-space QDs.

\section{Why Quantum Droplet Interferometry?}\label{IIIF}

While atomic bright solitons have been widely explored as one of the primary candidates for matterwave interferometry \cite{helm2014splitting,wales2020splitting,helm2012bright,helm2015sagnac}, it is worth comparing the merits of the soliton- and droplet-based interferometers. Quantum droplets exhibit some promising properties that make them most suitable for the use in the matter-wave interferometry: (i) nrondispersive nature, (ii) reduced phase-diffusion rate, and (iii) stability beyond the strict quasi-1D limit. To understand how the phase diffusion affects the interferometric performance, it is useful to consider the dephasing process induced by atomic interactions. Quantitatively, the relative phase accumulation rate between the two interferometer arms is set by the difference in their chemical potentials \cite{berrada2013integrated},
	\begin{equation}
		\dot{\phi}(t)=-\delta\mu, \qquad \delta\mu=\mu_L-\mu_R.
	\end{equation}
	The corresponding phase variance evolves as \cite{berrada2013integrated,javanainen1997phase}
	\begin{equation}
		(\Delta\phi(t))^2=(\Delta\phi(0))^2+R^2 t^2,
	\end{equation}
	where
	\begin{equation}
		R=\Delta n \left(\frac{\partial\mu}{\partial\mathcal{N}}\right)_{\mathcal{N}=N/2}
	\end{equation}
	is the phase-diffusion rate for a $50:50$ splitting. Here $\Delta n=\xi_N \sqrt{N}$ denotes the standard deviation of the atom number difference $n=N_L-N_R$ and $\xi_N$ is unity in the absence of squeezing.

In case of \emph{bright solitons}, the chemical potential is given by $\mu=-\frac{|g|^2 \mathcal{N}^2}{8}$ with $g=2a_{s}/l_{\perp}$, where $a_{s}$ is the $s$-wave scattering length and $l_{\perp}$ the transverse oscillator length \cite{carr2002dynamics,salasnich2017bright}. This result in a phase-diffusion rate of $R_{\mathrm{soliton}}=-\frac{|g|^2}{8} N^{3/2}$, which grows rapidly with atom number. On the contrary, for \emph{1D quantum droplets}, the rate of the change of chemical potential with respect to atom number decreases with the increase of atom number. Specifically, in the flat-top regime, the chemical potential saturates at $\mu=-2/(9\gamma)$, where $\gamma$ is the relative mean-field interaction strength (Eq.~(\ref{mu0})). Consequently, the phase-diffusion rate decreases with the increase of the atom number and effectively vanishes in the flat-top regime, making droplets highly favorable for the interferometry. \emph{Noninteracting matterwave packets} also exhibit a constant chemical potential \cite{pethick2008bose}, implying no phase diffusion. However, their dispersive nature causes spatial spreading during the interferometric cycle, reducing the fringe visibility. In contrast, the droplets combine the advantages of suppressed dispersion and phase diffusion at large atom numbers, making them uniquely well-suited for matter-wave interferometry.

In addition, the \emph{bright soliton} destabilize and eventually develop collapse or transverse instabilities once the transverse confinement is relaxed \cite{MalomedMS,salasnich2017bright}.	A rough crossover can be estimated by comparing the transverse confinement length scale $l_\perp$ with the characteristic size of the soliton or, equivalently, by comparing the chemical potential with the transverse confinement energy. The crossover regime is reached when these two energy scales are comparable. Using the 1D soliton expression for the chemical potential, $\mu=-\frac{|a_{s}N|^2}{2l^2_{\perp}}$ (in the units of $\hbar\omega_\perp$) \cite{carr2002dynamics,salasnich2017bright}, the crossover condition is approximately given by $|a_{s}|N=\sqrt{2}l_{\perp}$. This relation restricts the choice of number of atoms $N$ or s-wave scattering length $a_{s}$ for given $l_{\perp}$, defining the limit of the validity of the quasi-1D regime. When $|a_s|N>\sqrt{2}l_{\perp}$, the bright soliton collapses, wherease the droplets remain self-bound even in higher dimensions \cite{luo2021new}.

\begin{figure*}[th]
	\centering
	\includegraphics[width=16 cm]{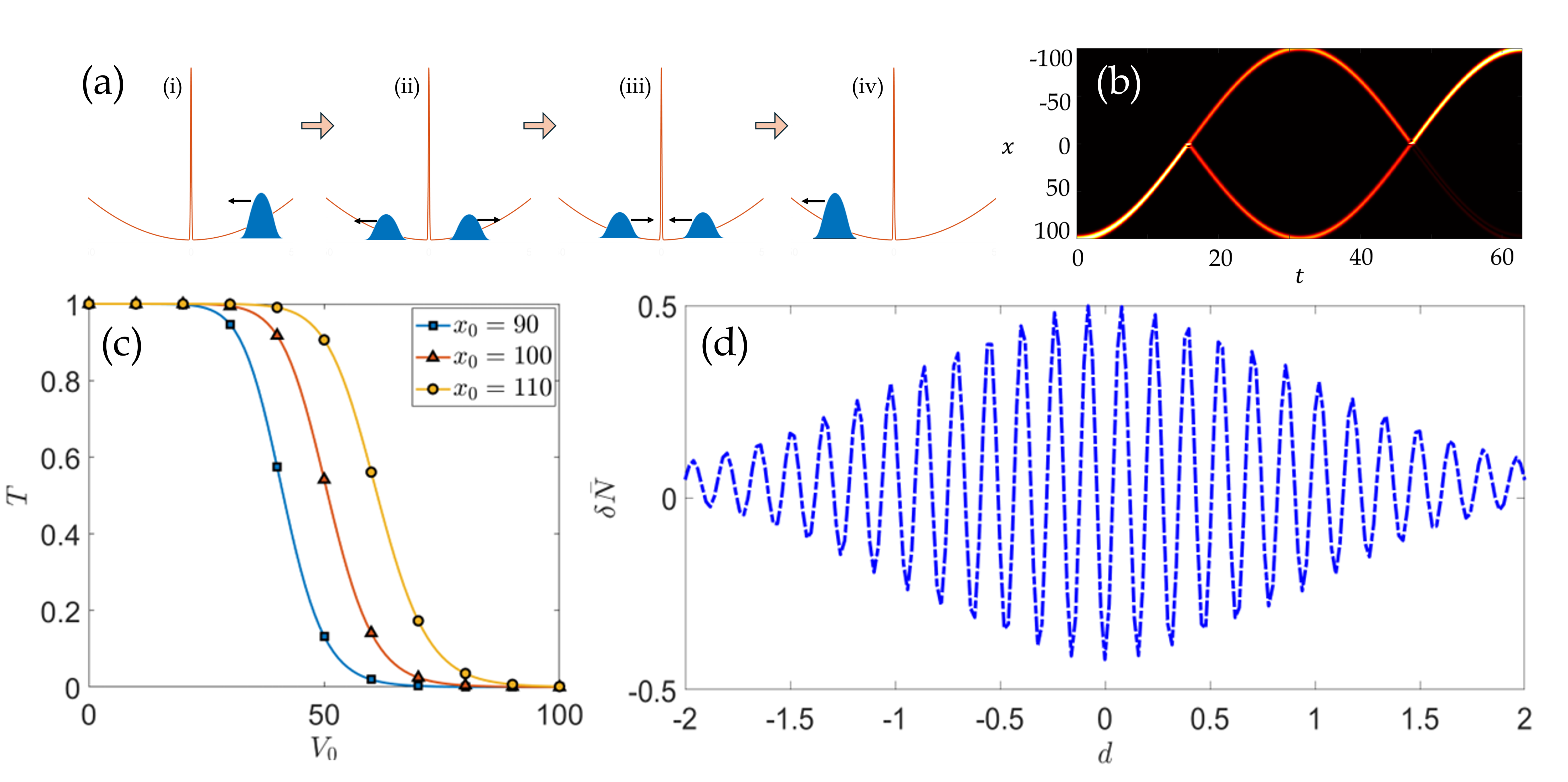}
	\caption{(a) The schematic of the QD interferometer based on the HO trap.
		(i) The prepared ground state is displaced by distance $x_{0}$ and released to move freely. (ii) Splitting of the moving QD by the central Gaussian barrier, (iii) and (iv) Recombination of the daughter droplets at the barrier. (b) The space-time trajectory of the interferometric sequence. (c) Dependences of the transmission coefficient on the barrier's height for different initial displacements of the QD, $x_{0}=90,\;100$, and $110$, are
		indicated by lines with squares, triangles and circles, respectively. (d) The sensitivity of the relative atom number, past the recombination, to the barrier's position. Here, the frequency of the harmonic trap is $\protect%
		\omega _{x}=0.1\protect\omega _{\perp }$ and the width of the Gaussian barrier is $d_{0}=0.25$. The droplet is prepared with $\protect\gamma =3$ and $N=0.5$. Coordinate $x$ and time $t$ are measured in units of $l_{\perp}=0.72\;\mathrm{\protect\mu }$m and $1/\protect\omega _{\perp }=0.32\;$ms, respectively.}
	\label{Sch_HQD}
\end{figure*}
\section{The QD interferometers}
\label{SecIII}

\subsection{The barrier interferometry with HO-trapped QDs}

\label{IIIA} To initiate the operation of the interferometry scheme with
HO-trapped QDs, we numerically prepared a ground-state QD, which is then displaced by distance $x_{0}$ from the trap's center and released
to move, colliding with a narrow potential barrier placed at the center. A
schematic of the respective setup is presented in Fig.~\ref{Sch_HQD}(a). It
involves the initialization and $50:50$ splitting of the QD (stages (i) and
(ii)), followed by the recombination of the daughter (secondary) droplets at
the barrier (stages (iii) and (iv)). The operation was modeled, using
real-time simulations \cite{bera2020matter} of Eq.~(\ref{1DQDGPE}), which
includes the potential combining the HO trap and central splitting barrier:
\begin{equation}
	V(x)=\frac{1}{2}\lambda ^{2}x^{2}+V_{\mathrm{barrier}}(x).  \label{1DHQD_I}
\end{equation}%
The barrier is modeled by a Gaussian with width $d_{0} $ and integral
strength $q\equiv \int_{-\infty }^{+\infty }V_{\mathrm{barrier}}(x)\mathit{d}x$,
that determines the barrier height, $V_{0}=q/\left( \sqrt{2\pi }d_{0}
\right) $:
\begin{equation}
	V_{\mathrm{barrier}}(x)=V_{0}\exp \Big(-\frac{x^{2}}{2 d_{0} ^{2}}\Big).
	\label{Gbarrier1}
\end{equation}%
In the experiment, the Gaussian barrier can be readily imposed by a blue-detuned laser beam \cite{marchant2013controlled}. The barrier is
sufficiently narrow, making it possible to neglect velocity-filtering
effects, which refer to the energy-dependent transmission and reflection of
different QD's velocity components (so that faster components are more
likely to be transmitted, while slower ones are preferentially reflected \cite{polo2013soliton,wales2020splitting,cuevas2013interactions}).
Figure \ref{Sch_HQD}(b) displays the overall space-time plot (trajectory) of the operation, which includes the splitting and recombination of the QD.
To identify the barrier height securing the $50:50$ splitting, we
numerically computed the transmission coefficient, defined as:
\begin{equation}
	T=\frac{\int_{-\infty }^{0}|\psi (x,t=t_{1})|^{2}\mathit{d}x}{\int_{-\infty
		}^{+\infty }|\psi (x,t=0)|^{2}\mathit{d}x}.  \label{T}
\end{equation}%
As the initial droplet is displaced to $x_{0}=100$, the right-hand side of the barrier, Eq. (\ref{T}) defines the transmission coefficients by the
integration over the region $x<0$, while  $t_{1}$ is the instant of time at which the droplet has fully split into two daughter droplets, which moved to positions far from the barrier. Here, the barrier is narrow and $t_{1}$ is long enough that, any transiently trapped component is negligible, and the two-channel picture ($R+T=1$) is accurate to numerical precision. However, for wide barriers (compared to QD width), an additional “trapped” fraction should be included \cite{konotop1996interaction,damgaard2021scattering}. For a fixed barrier width, $T$ depends on the barrier height, as shown in Fig.~\ref{Sch_HQD}(c). $T$ exhibits a
characteristic sigmoid response: it is $\approx 1$ for a low barrier, gradually approaching $0$ for higher $V_{0}$. The barrier height required for the $50:50$ transmission is affected by the velocity of the QD when it is
colliding with the barrier, which, in turn, is determined by the initial
displacement $x_{0}$. Naturally, larger $x_{0}$ results in a higher collision velocity, requiring a greater barrier height to achieve $T=0.5$.  As shown in the previous section, at large $%
\gamma$ the crossover from the QDs to HO-determined bound states occurs at smaller $N$, highlighting the need to carefully select parameters for the realisation of the QD interferometry. Thus, a key requirement for the successful realization of QD interferometry based on HO trap is to ensure  $\gamma < \gamma_{\mathrm{thr}}$, as exceeding this threshold eliminates the QD regime.

The daughter droplets produced by the $50:50$ splitting in the HO trap
return to the location of the splitting barrier, where the collision between
them leads to the interference-driven recombination. The final atomic
population on either side of the barrier after the recombination depends on
the phase difference between the daughter droplets. The definitions of the
numbers of atoms on the right (positive) and left (negative) sides of the
barrier are straightforward:
\begin{equation}
	N_{\pm }=\pm \int_{0}^{\pm \infty }|\psi (x,t_{2})|^{2}\mathit{d}x
	\label{HQD2BN}
\end{equation}%
cf, Eq.~(\ref{T}). Here, $t_{2}$  is the time moment at which the initial
droplet has completed its cycle of splitting, recombination, and separation
from the barrier. In the case of the delta-functional barrier, the splitting produces the phase difference of $\pi /2$ between the \textquotedblleft daughters" \cite{helm2012bright,wales2020splitting}. This feature results in a scenario where the entire population is transferred to the left-hand side of the barrier ($N_{-}=1$). However, in realistic scenarios, factors such as the finite barrier width and the internal nonlinearity of the droplet cause
deviations from the ideal $\pi /2$ phase difference. As we consider a narrow
barrier, the phase difference between the daughter droplets is primarily
determined by the relative interaction strength $\gamma $.

The barrier interferometry is based on the fact that the population on
either side of the barrier after the recombination depends on the phase
difference between the daughter droplets at the time of the collision
between them. The phase difference can be controlled by adjusting the
position of the narrow barrier. A shift in the barrier position alters the
paths of the transmitted and reflected \textquotedblleft daughters",
introducing a spatial displacement between them. The displacement leads to a
velocity-induced phase gradient across the droplets during the
recombination, which, in turn, affects the atom population on either side of
the barrier. The relative atom-number population is defined as $\delta \bar{N}=\left( N_{+}-N_{-}\right) /\left( 2N\right) $, so that values $\delta \bar{N}=+0.5$ and $-0.5$ imply that all the atoms are found, severally, on the right- and left-hand side of the barrier \cite{martin2012quantum}. Figure \ref{Sch_HQD}(d) displays the dependence of the relative atom number on the
offset of the barrier position in the case of $\gamma =3$, $N=0.5$. The
observed oscillations of the relative atom number characterise the
sensitivity of the interferometer with respect to the barrier position.
Because the phase imprint of $\pi $ induces a half-cycle oscillation in $\delta \bar{N}$ \cite{martin2012quantum,helm2012bright}, the barrier offset
by $\approx 60\;$nm corresponds to the phase shift of $\pi $. The envelope
modulating the fringes is produced by the varying spatiotemporal overlap
between the daughter droplets \cite{wales2020splitting}.
\begin{figure*}[th]
	\centering
	\includegraphics[width=16 cm]{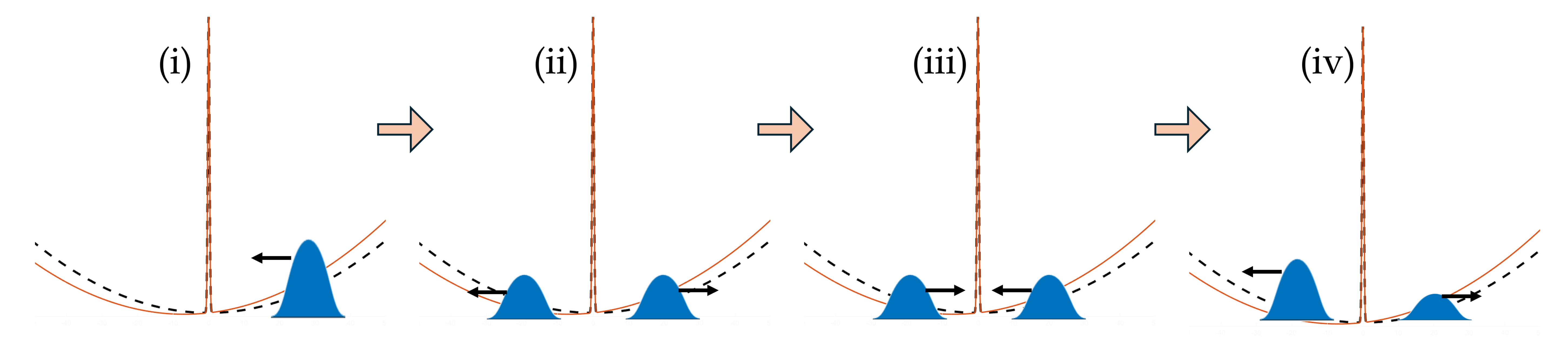}
	\caption{The schematic of the tilt-meter using the QD in the HO trap. The
		trap is tilted by angle $\protect\theta $, while the barrier is centered at $x=0$. The red and dashed black lines indicate the tilted and untilted HO\ trap, respectively. (i) and (ii):  The splitting of the initial QD in two daughter droplets, (iii) and (iv): The (incomplete) recombination of the daughter droplets at the barrier.}
	\label{HQDI_method}
\end{figure*}

\subsection{Tilt-metry with the HO-trapped QD interferometer}\label{IIIB} 
\begin{figure*}[th]
	\centering
	\includegraphics[width=16 cm]{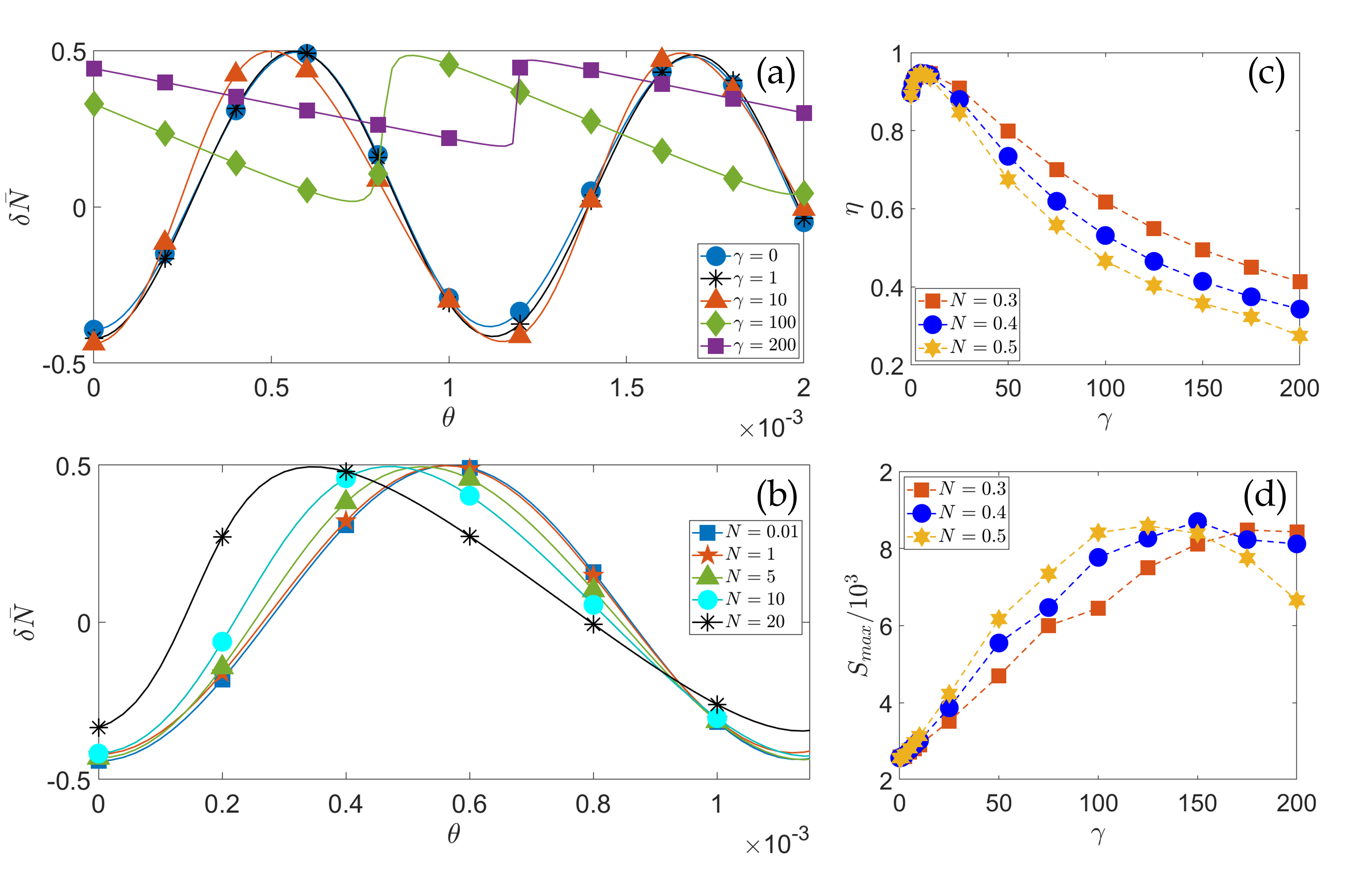}
	\caption{(a) Ocillations of the relative atom number with respect to 
		the tilt angle for different values of $\protect\gamma $. Here, the total atom number is taken as $N=0.5$ (b) The same oscillations with respect to\ the tilt angle for different values of $N$ where $\gamma=1$. (c) The variation of the contrast of the relative atom-number oscillations, defined as per Eq.(\protect\ref{Perf}), with respect to $\protect\gamma $ for different values of $N$. (d) The dependence of the maximum sensitivity of the tilt-meter on $\protect\gamma $, defined as per Eq. (\protect\ref{Perf}), for different values of $N$. Here, the HO-trap frequency is $\protect\omega _{x}=0.1\protect\omega _{\perp }$, and the width of the Gaussian barrier is $d_{0}=0.25$; $N=1$ corresponds to $1.8\times 10^{4}$ atoms.}
	\label{HQDI_result}
\end{figure*}

In the previous subsection, we discussed how the change in the
barrier position produces relative atom-number oscillations in the present
interferometer setup. The sensitivity of the interferometer with respect to
the barrier position can be naturally used for the design of sensing
applications. In this subsection, we report an example of that, addressing effects of the tilt added to the HO trap on the interferometric output. We introduce the tilt by adding a linear potential to Eq.~(\ref{1DHQD_I}):
\begin{equation}
	V(x)=\frac{1}{2}\lambda ^{2}x^{2}+\bar{g}x\tan {\theta }+V_{\mathrm{barrier}%
	}(x).  \label{tiltpot}
\end{equation}%
Here, $\bar{g}=g/\left( l_{\perp }\omega _{\perp }^{2}\right) $, where $g$
is the original tilt (in particular, it may be the acceleration induced by
gravity), and $\theta $ is the tilt angle. Figure \ref{HQDI_method} presents
a schematic of the respective QD-based tilt-meter, showing the HO trap in the
tilted (red) and untilted (dashed black) configurations. With the barrier
position fixed, tilting the trap induces relative atom-number oscillations,
akin to those displayed in Fig.~\ref{Sch_HQD}(d). We analyzed these
oscillations as a function of $\theta $ for different QD parameters $\gamma $
and $N$. For fixed $N=0.5$, the relative atom number $\delta \bar{N}$ exhibits the sinusoidal behavior at small $\gamma$ (for $\gamma<10$). It is observed that a phase shift of $\pi $ is obtained for a small tilt angle $\theta =5\times 10^{-4}\;$rad. At large $\gamma $ (e.g., $50$ or $100$), the response takes the form of a sawtooth-like pattern, as shown in Fig.~\ref{HQDI_result}(a). For given $\gamma=1$, varying the atom number $N$ of the initial QD produces a similar effect, as shown in Fig.~\ref{HQDI_result}(b). The atom number oscillations start out as a sinusoidal oscillations with respect to $\theta$ and gradually becomes a sawtooth oscillations with the  increase of $N$ or $\gamma$. 

The performance of the tilt-meter is quantified by contrast $\eta $ and
maximum sensitivity $S_{\max }$ as follows:
\begin{equation}
	\eta =\mathrm{max}(\delta \bar{N})-\mathrm{min}(\delta \bar{N}),\;\;S_{\max }=%
	\mathrm{max}\Big(\Big|\frac{\mathit{d}}{\mathit{d}\theta }\delta \bar{N}\Big|%
	\Big).  \label{Perf}
\end{equation}%
An efficient interferometer is expected to exhibit a high value of $S_{\max }
$ while maintaining a contrast close to $1$. Figures \ref{HQDI_result}(c)
and (d) depict the variation of $\eta $ and $S_{\max }$ as functions of $%
\gamma $ for different atom numbers $N$. At small values of $\gamma $, the
atom number has a little effect on the contrast and maximum sensitivity.
However, at large values of $\gamma $, the increase of $N$ leads to a
decrease in the contrast. High contrast is observed near the QD-HO determined bound states crossover boundary, followed by a steady decline in the contrast as $\gamma$ increases, regardless of the value of $N$. On the other hand, $S_{\max }$ increases with $\gamma $ and then decreases after reaching a maximum at a particular value of $\gamma $. This value decreases with the increase of $N$. For instance, $\gamma =100$ yields very good sensitivity at $N=0.5$, but with a low contrast of $0.5$. Therefore, it is necessary to choose values of $\gamma $ and $N$, which uphold good contrast and sensitivity without a significant compromise.
\begin{figure}[th]
	\centering
	\includegraphics[width=10 cm]{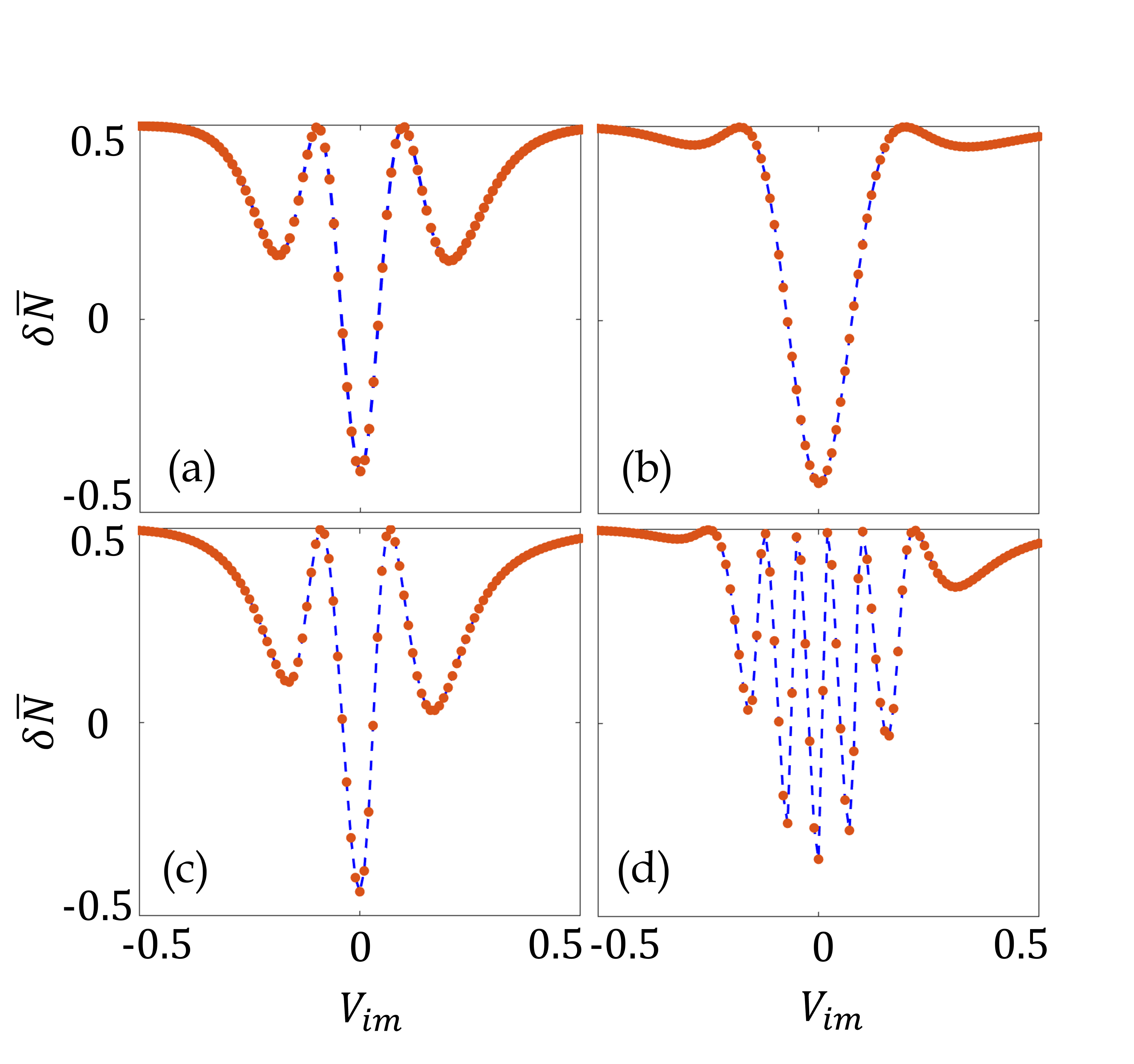}
	\caption{The change of the relative atom number after the recombination,
		following the variation of the impurity's height for different $\protect\gamma $: (a) $\protect\gamma =0$, (b) $\protect\gamma =1$, (c) $\protect\gamma =10$, (d) $\protect\gamma =25$. The total atom number is taken as $N=0.5$. The frequency of the HO trap is $\protect\omega _{x}=0.1\protect\omega _{\perp }$, and the widths of the Gaussian barrier and impurity are $d_{0}=d_{im}=0.25$. The height of the impurity $V_{im}$ is scaled by the height of the central barrier $V_{0}$.}
	\label{Loaded_Inf}
\end{figure}
\begin{figure}[th]
	\centering
	\includegraphics[width=15 cm]{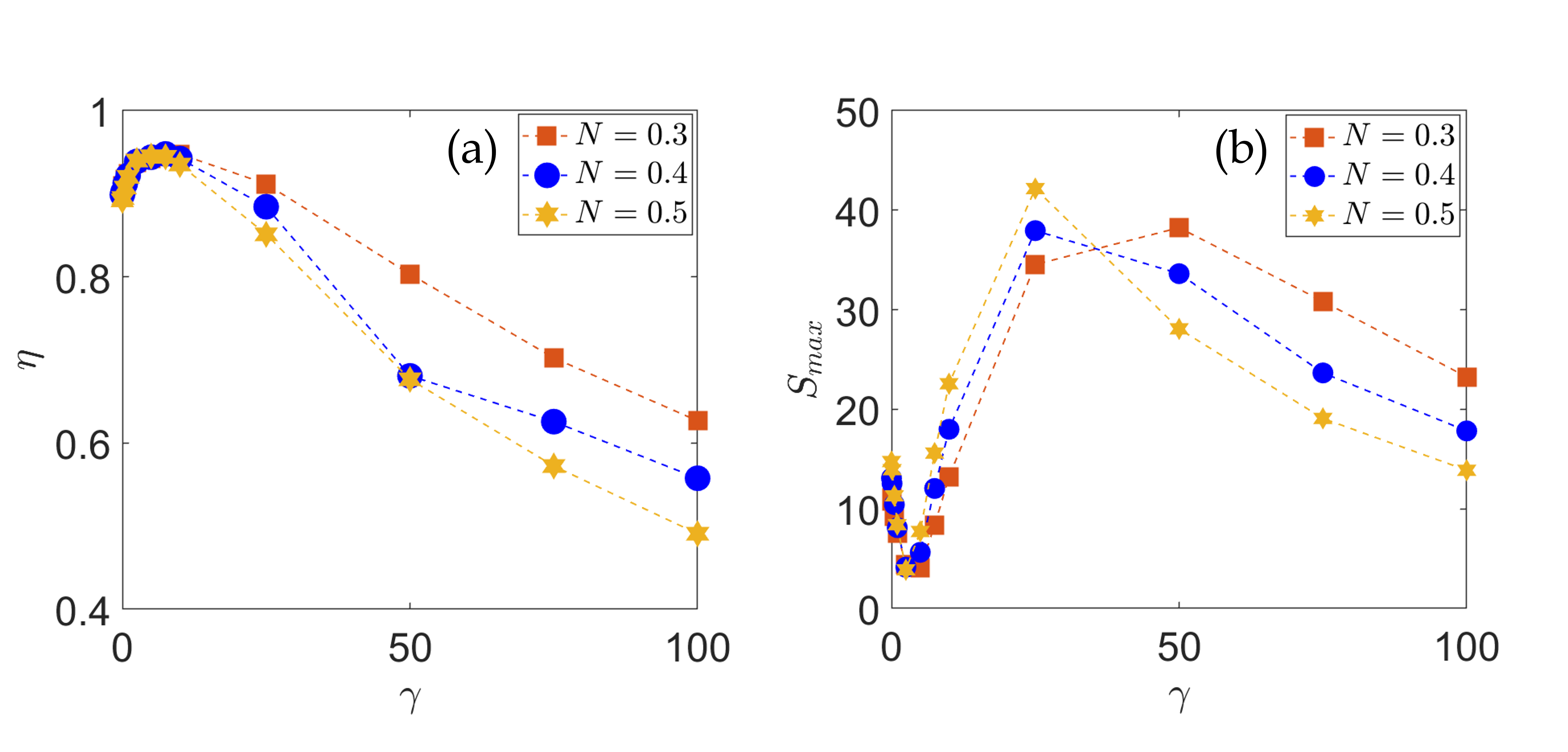}
	\caption{(a) The variation of the contrast of $\delta\bar{N}$,
		defined as per Eq.(\protect\ref{Perf}), with respect to $\protect\gamma $ for different values of $N$. (b) The dependence of the maximum sensitivity of target detection on $\protect\gamma $, defined as per Eq. (\protect\ref{Perf}), for different values of $N$. Here the frequency of the harmonic trap is $\omega_x=0.1\omega_{\perp}$ and width of the Gaussian barrier and impurity is taken as $d_{0}=d_{im}=0.25$. The height of the impurity $V_{im}$ is scaled by the height of the central barrier $V_0$.}
	\label{Loaded_Smax}
\end{figure}
\subsection{Target detection with HO-trapped QD interferometer}\label{IIIC}
Another important application of the HO-trapped interferometer is in target detection, when one arm of the interferometer is loaded with a local target \cite{sakaguchi2016matter}. The target is modelled as a Gaussian impurity with width $d_{im}$ and height $V_{im}$. For the definiteness' sake, we set the width of the impurity equal to that of the central barrier. Interferometric studies are then performed by varying the impurity height in the range $[-V_0/2, V_0/2]$, where $V_0$ is the height of the central barrier. The target is placed at $-x_0/2$ to the left of the origin so that the transmitted droplet interacts with it twice. This interaction induces a phase shift in the droplet, leading to incomplete recombination of the transmitted and reflected droplets at the barrier. Figure \ref{Loaded_Inf} shows the change of the relative atom number $\delta\bar{N}$ following variation of target height $V_{im}$ for different values of relative MF nonlinearity strength: (a) $\gamma=0$, (b) $\gamma=1$, (c) $\gamma=10$ and (d) $\gamma=25$. The total atom number is taken as $N=0.5$. The response of the interferometer to the repulsive and attractive target is almost identical for the range of target height considered in our study. For $\gamma=25$, a target of height $V_{im}=\pm V_0/50$ imposes a phase shift of $\pi$. 

\begin{figure*}[th]
	\centering
	\includegraphics[width=16 cm]{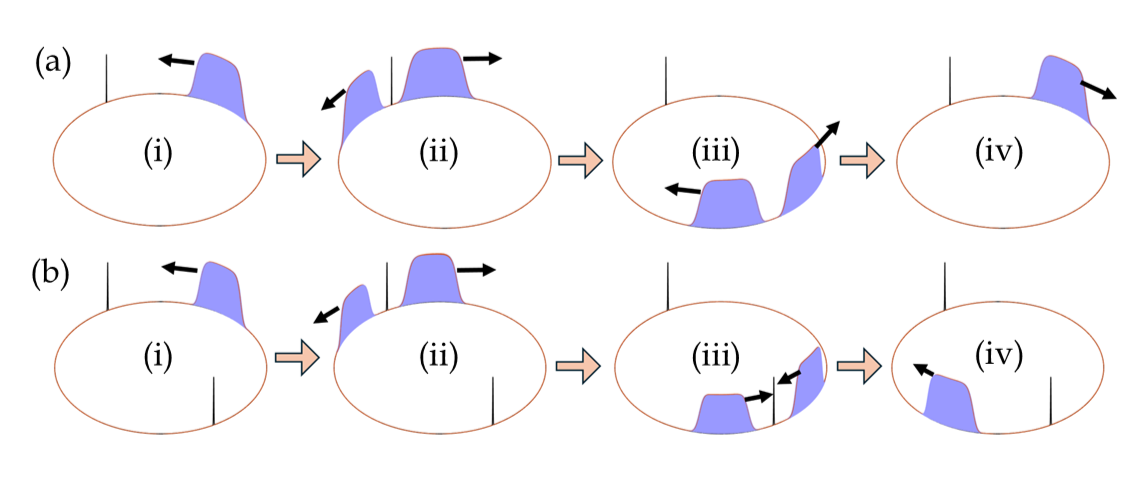}
	\caption{The schematic of the QD interferometer based on the ring trap with (a) a single and (b) double barrier: (i) and (ii): The splitting of the initial QD in identical daughter droplets. (iii) and (iv): The
	recombination of the daughter droplets at the barrier.}
	\label{RQDI_Scheme}
\end{figure*}
The quality of target detection is characterized using the contrast $\eta$ and the maximum sensitivity $S_{\mathrm{max}}$, as defined in Eq.~(\ref{Perf}). In Fig.~\ref{Loaded_Smax}, we show the dependence of the contrast in relative atom number on $\gamma$ for different values of the total atom numbers. High contrast is observed near the crossover boundary between the QDs and HO-determined bound states, similar to the behavior reported above in the tilt-meter case. For $\gamma > 5$, the contrast steadily decreases regardless of atom number. The maximum sensitivity $S_{\mathrm{max}}$ exhibits a minimum near the crossover boundary and increases to a peak at $\gamma \approx 25$. Beyond this point, $S_{\mathrm{max}}$ decreases with increase of $\gamma$. To achieve high sensitivity and contrast of up to $90\%$, operating around $\gamma = 25$ is optimal.

\subsection{The barrier interferometry with QDs in a ring trap}

\label{IIID} To analyze the operation of the QD-based interferometer in the
ring configuration, we consider Eq.~(\ref{1DQDGPE}) with periodic boundary
conditions at $x\in \lbrack -\pi r_{0},\pi r_{0}]$, where $r_{0}$ is the
radius of the ring. This ring geometry allows us to explore two different
scenarios: (1) a single barrier and (2) a pair of diametrically opposite ones:
\begin{equation}
	V_{\mathrm{barrier}}(x)=V_{0}\Big[\exp \Big(-\frac{x^{2}}{2 d_{0} ^{2}}\Big)+%
	\mathrm{b}\exp \Big(-\frac{(x\pm \pi r_{0})^{2}}{2 d_{0} ^{2}}\Big)\Big].
\end{equation}%
Here, $V_{0}=q/\left( \sqrt{2\pi }d_{0} \right) $ is the barrier height as
above, while $b=0$ or $1$ corresponds to the single barrier or the pair of
them. Due to the Galilean invariance of Eq.~(\ref{1DQDGPE}) (far from the
barrier(s)), we take the input as
\begin{equation}
	\psi (x,0)=-\frac{3\mu e^{-ivx}}{1+\sqrt{1+9\mu \gamma /2}\cosh \left( \sqrt{%
			-2\mu }(x-x_{0})\right) },
	\label{input}
\end{equation}%
where $v$ is the QD's velocity in units of $l_{\perp }\omega _{\perp }$. The
initial velocity can be experimentally implemented through the momentum
transfer imparted by Bragg transitions \cite%
{kozuma1999coherent,ernst2010probing}.

Unlike the HO-trapped QDs, at high values of $\gamma $,
there is no restriction on the value of $N$ for the existence of self-bound
QDs, as we take the free-space solution Eq.~(\ref{input}) as the input. An important remark is that the QD's width must be less than half the circumference of the ring, to let the split daughter droplets remain distinguishable and separated. The latter condition is crucial for accurately measuring the relative phase of the \textquotedblleft daughter", which, in turn, is essential for high-precision sensing applications, such as rotation measurements. In the flat-top regime, the width increases
linearly with $N$ and quadratically with $\gamma $.
\begin{figure}[th]
	\centering
	\includegraphics[width=8 cm]{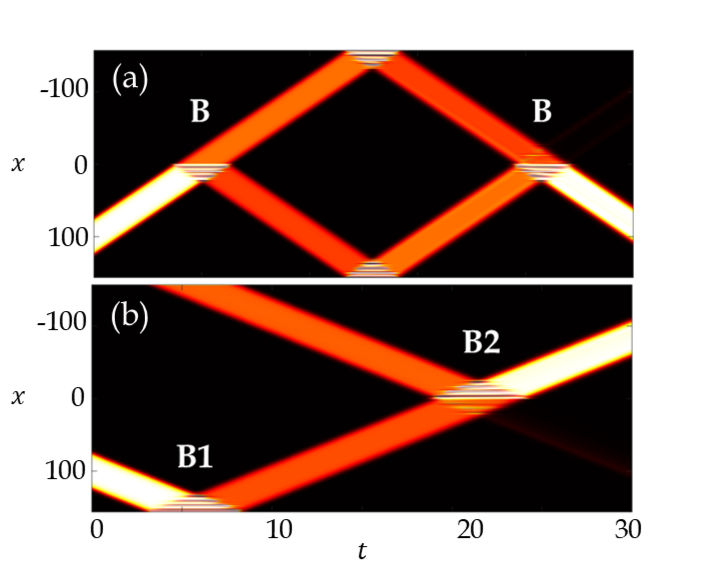}
	\caption{The spatiotemporal plot of the operation of the ring-shaped
		interferometer with (a) the single barrier (B) and (b) the pair of diametrically opposite ones (B1,B2). Here, the radius of the ring is $r_{0}=60$, and the width of the Gaussian barrier is $d_{0}=0.25$. $x$ and $t$ are measured in units of $l_{\perp}=0.72\;\mathrm{\protect\mu }$m and $1/\protect\omega _{\perp }=0.32\;$ms, respectively.}
	\label{RQDI_ST}
\end{figure}

A scheme of the ring-based QD interferometer is drawn in Fig.~\ref{RQDI_Scheme}, where panels (a) and (b) represents the single- and
double-barrier configurations, respectively. The stages of the
interferometry operation are similar to those discussed above for the
HO-trap setup: panel (i) and (ii) outline the initialization and splitting,
while (iii) and (iv) depict the recombination of the daughter droplets at
the barrier. The space-time plots of the interferometer operation,
corresponding to the single- and double-barrier schemes are presented in
Figs.~\ref{RQDI_ST}(a) and (b), respectively.

For the pair of diametrically opposite narrow barriers (at $x=\pm \pi r_{0}$ and $x=0$), the initial droplet, placed at $x_{0}=100$ with positive velocity $v$, interacts with the first barrier, fixed at $x=\pi r_{0}$, splitting into pair of daughter droplets. Here, we consider the initial QD with $\gamma =3$ and $N=2.3$. We determine the barrier height that corresponds to the $50:50$ splitting, using the transmission coefficient defined as
\begin{equation}
	T=\frac{\int_{-\pi r_{0}}^{0}|\psi (x,t_{1})|^{2}\mathit{d}x}{\int_{-\pi
			r_{0}}^{+\pi r_{0}}|\psi (x,0)|^{2}\mathit{d}x},  \label{T2}
\end{equation}%
\begin{figure*}[th]
	\centering
	\includegraphics[width=16 cm]{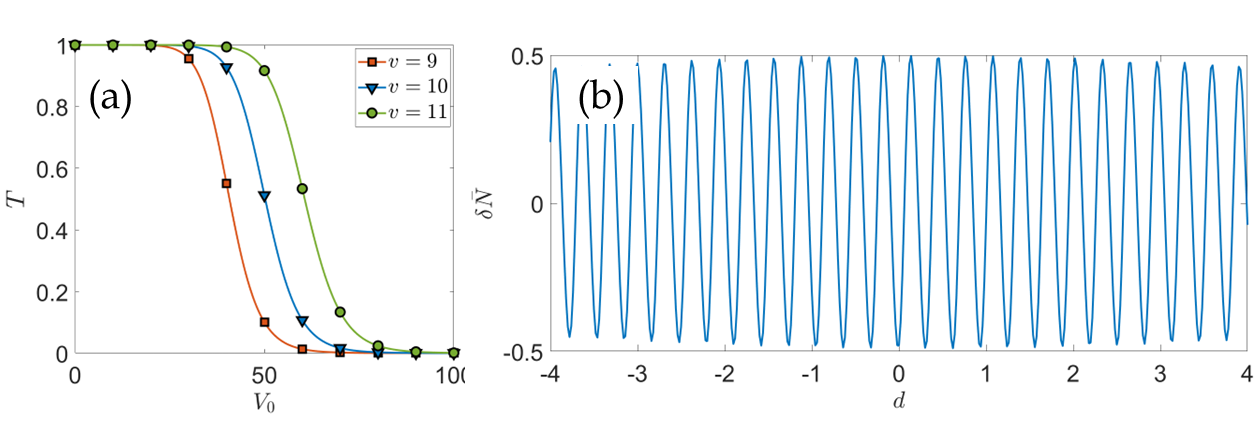}
	\caption{(a) The dependence of the transmission coefficient in the ring-shaped interferometer on the barrier height for different initial velocities of the QD, $v=9,\;10$, and $11$, is indicated by lines with squares, triangles and circles, respectively. (b) The sensitivity of the relative atom number after the recombination at the position of the second barrier. The radius of the ring is $r_{0}=60$, the two barriers are placed at diameterically opposite points, \textit{viz}., $x=\protect\pi r_{0}$ and $x=0$. The width of the Gaussian barrier is $d_{0}=0.25$. Here, the droplet is prepared for $\gamma=3$ and $N=2.3$. $x$ and $t$ are measured in units of $l_{\perp }=0.72\;\mathrm{\protect\mu }$m and, $1/\protect\omega _{\perp }=0.32\;$ms, respectively.}
	\label{RQDI_TC}
\end{figure*}
cf. Eq. (\ref{T}). Here $t_{1}$ is time at which the initial QD has fully split in the pair of \textquotedblleft daughters", that are located sufficiently far from the barrier. Figure \ref{RQDI_TC}(a) shows the dependence of the transmission coefficient on the barrier's height for different initial velocities $v$. A sigmoid curve is observed, similar to the HO-trap case, as seen in Fig.~\ref{Sch_HQD}(c), where faster initial QDs require greater barrier heights to achieve $T=$ $0.5$. Upon the splitting, the daughter droplets move towards the opposite barrier at $x=0$, where their overlap leads to\ the interference-driven recombination. The resultant population of atoms on either side of the second barrier depends on the phase difference of the \textquotedblleft daughters". The populations
on the either sides of the second barrier, $N_{+}$ and $N_{-}$, are defined as:
\begin{equation}
	N_{\pm }=\pm \int_{0}^{\pm \pi r_{0}}|\psi (x,t_{2})|^{2}\mathit{d}x,
	\label{RQD2BN}
\end{equation}%
cf. Eq. (\ref{HQD2BN}). Here, $t_{2}$ is an instant of time at which the recombined QD is far separated from the barrier. This definition of the populations is relevant for both the single- and two-barrier cases, as in both cases the recombination happens at $x=0$. The sensitivity of the interferometer with respect to the barrier position is studied, offsetting the position of the second barrier from $x=0$ by a shift $d$ and observing the resultant population on both sides of the second barrier. For this study, we define the relative atom number in the same way as above, $\delta \bar{N}=\left(N_{+}-N_{-}\right) /\left( 2N\right) $. Relative-atom-number oscillations due to the change in barrier position are displayed in Fig.~\ref{RQDI_TC}(b). In this case, a phase shift of $\pi $ is brought about by the barrier offset $\approx 100\;$nm.
\begin{figure*}[th]
	\centering
	\includegraphics[width=16 cm]{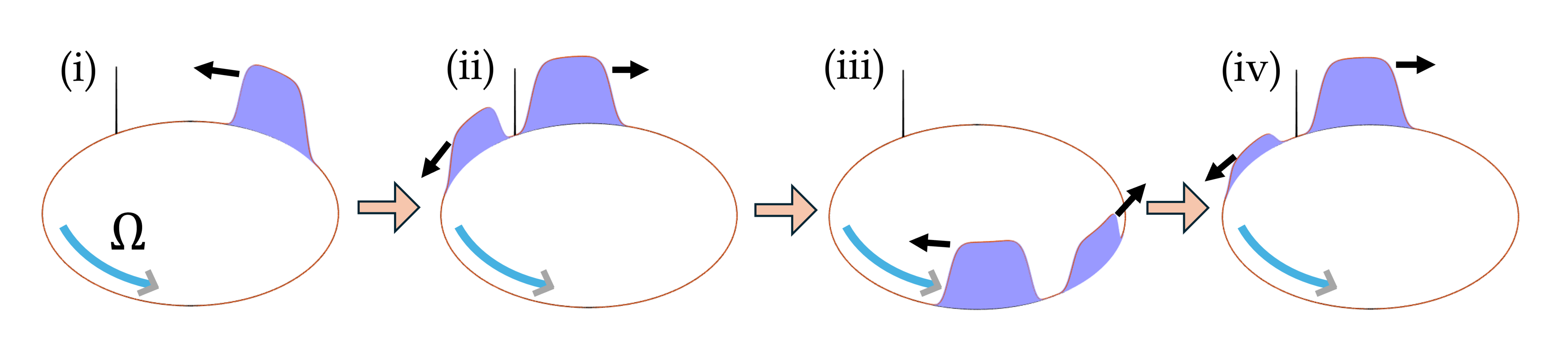}
	\caption{The schematic of Sagnac interferometer using QD in a ring: (i) and (ii) The splitting of the initial QD into two identical daughter droplets, (iii) and (iv) The recombination of the daughter droplets at the barrier. Here, $\Omega $ denotes the rotation frequency of the ring.}
	\label{RQDI_method}
\end{figure*}

\subsection{Sagnac interferometry with QDs} \label{IIIE} 
\begin{figure*}[th]
	\centering
	\includegraphics[width=16 cm]{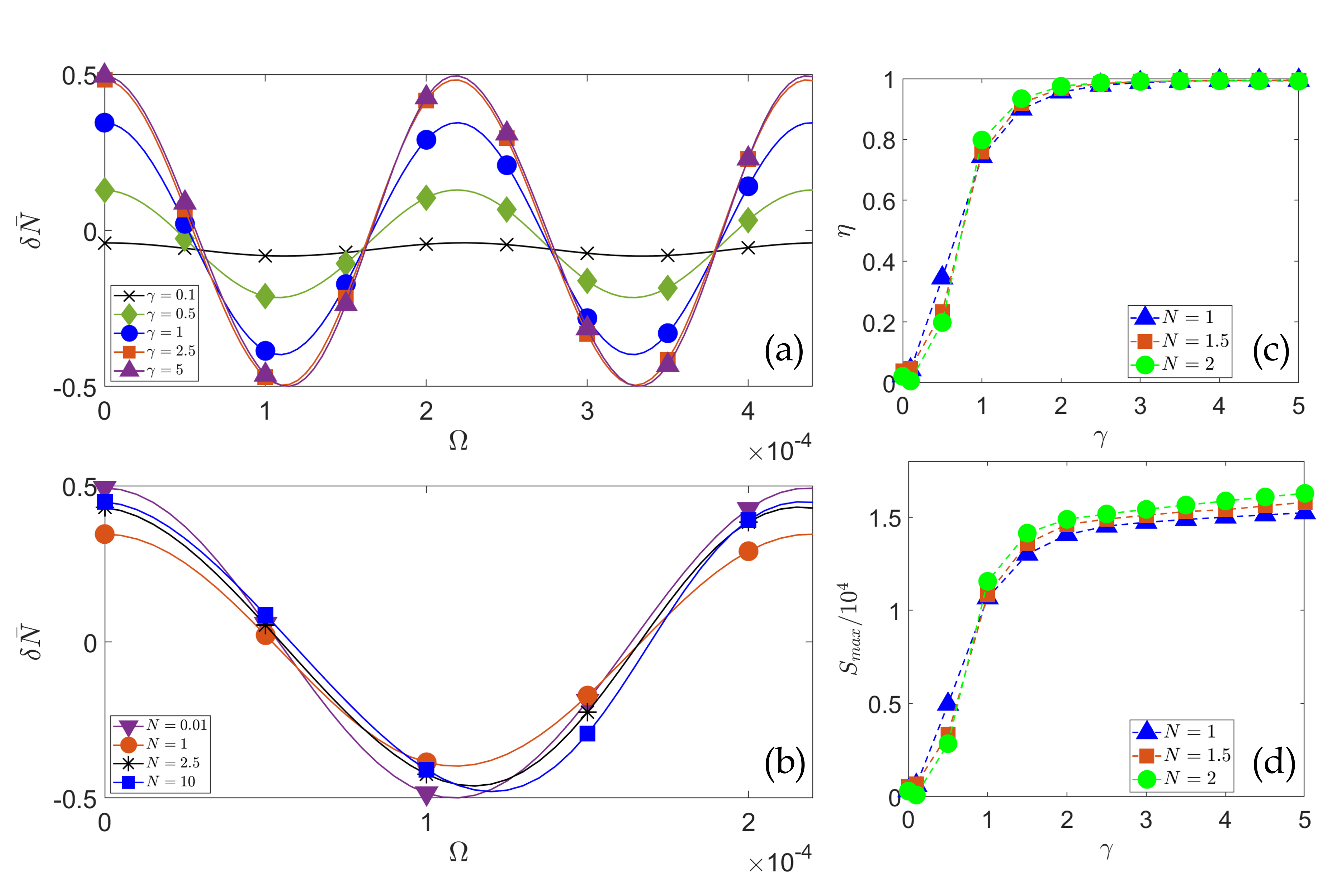}
	\caption{(a) Relative atom number oscillation with respect to rotation
		frequency for different $\protect\gamma $. Here, the total atom number considered is $N=1$ (b) Relative atom number
		oscillation with respect to rotation frequency for different values of $N$, with $\gamma=1$. (c) The variation of contrast of the relative atom number oscillation with respect
		to $\protect\gamma $ for different values of $N$. (d) The dependence of maximum sensitivity of the tilt-meter with $\protect\gamma $ for different values of $N$. The radius of the ring is taken as $r_{0}=60$ and width of the Gaussian barrier is $d_{0}=0.25$. Here, $N=1$ corresponds to $1.8\times 10^{4}$ atoms and $\Omega $ is in the unit of $\protect\omega _{\perp }$.}
	\label{RQDI_results}
\end{figure*}

The ring-based QD interferometer creates an additional phase
shift between the daughter droplets in a rotating interferometer. This shift
is the Sagnac phase, being proportional to the area $\mathbf{A}$ enclosed by
the ring and rotation frequency $\boldsymbol{\Omega }$ \cite{helm2015sagnac,burke2009scalable,gautier2022accurate}. In this context, we consider the 1D GPE in the rotating reference frame:
\begin{eqnarray}
	i\frac{\partial \psi (x,t)}{\partial t}=\Big[-\frac{1}{2}\frac{\partial ^{2}}{\partial x^{2}}&+&i\Gamma \frac{\partial }{\partial x}+\gamma |\psi(x,t)|^{2}  \nonumber \\
	&& \quad -|\psi (x,t)|+V_{\mathrm{barrier}}(x)\Big]\psi (x,t).  \label{RFQDGPE}
\end{eqnarray}%
where $\Gamma \equiv \Omega r_{0}$ is the linear velocity of the frame.
Because the Sagnac interferometer with single barrier is twice as sensitive
as the one with two barriers, we here use the setup with the single barrier.
\begin{figure}[ht]
	\centering
	\includegraphics[width = 17cm]{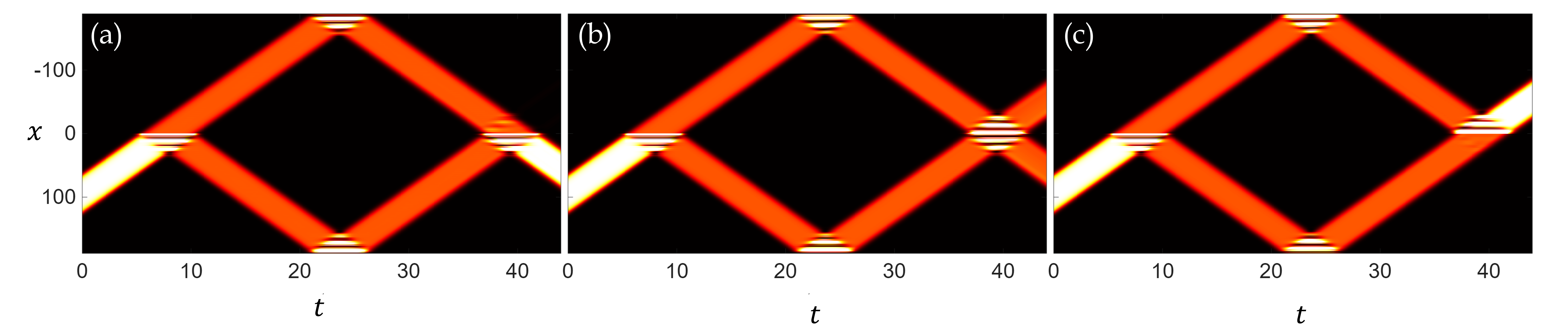}
	\caption{Spatiotemporal plot of quantum droplet density in a rotating Sagnac Interferometer for rotation frequency (a) $\Omega=0$, (b) $\Omega=0.57\times10^{-4}$ and (c) $\Omega=1.14\times10^{-4}$. The radius of the ring is taken as $r_0=60$, the width of the barrier as $d_0=0.25$ and the total atom number $N=1$. Here, $N=1$ corresponds to $1.8\times10^4$ atoms and $\Omega$ is in the units of $\omega_{\perp}$.}
	\label{STSI}
\end{figure}
Figure \ref{RQDI_method} outlines the operation of the Sagnac QD-based
interferometer. As mentioned above, the rotation frequency $\Omega $ adds
the Sagnac phase to the phase shift imposed by the splitting. The output of the Sagnac interferometer is sensitive to the rate at which the ring rotates, as shown in Fig.~\ref{STSI}. The spatiotemporal dynamics of the QD for rotation frequencies $\Omega = 0$, $\Omega = 0.57 \times 10^{-4}$, and $\Omega = 1.14 \times 10^{-4}$ are presented in Figs.~\ref{STSI}(a–c). For different
rotation frequencies, we calculated the relative atom number $\delta\bar{N}$, similar to the above analysis reported for the HO-trap interferometer. In Fig.~\ref{RQDI_results}(a), the dependence of the relative atom number on the rotation frequency is plotted for different values of $\gamma$ and fixed atom number, $N=1$. At high values of $\gamma $, the amplitude of the $\delta\bar{N}$ oscillation is the largest. In this case, the phase shift of $\pi$ is imposed for the rotation frequency of $\approx 10^{-4}\omega _{\perp }$. Similarly, keeping $\gamma=1$, Fig.~\ref{RQDI_results}(b) shows oscillations of the relative atom number with respect to $\Omega $, for different values of $N$ in the initial QD.

We quantify the performance of the Sagnac interferometer, using the
contrast, $\eta $, and maximum sensitivity, $S_{\max }$, defined as per Eq.~(
\ref{Perf}). The dependence of $\eta $ and $S_{\max }$ on $\gamma $ for
different values of $N$ is plotted in Figs.~\ref{RQDI_results}(c) and (d).
The rotation sensor exhibits low contrast and sensitivity in the
LHY-superfluid regime, corresponding to small values of $\gamma $. However,
both the contrast and sensitivity steadily improve with the increase of $%
\gamma $. The contrast reaches the maximum close to $1$ around $\gamma =2$
and then saturates, while the maximum sensitivity continues to increase
gradually beyond this point.

The increase of the sensitivity with $\gamma $ is a noteworthy finding,
because, as discussed earlier, at higher values of $\gamma $ the QD becomes somewhat
similar to the non-interacting BEC (Eq.~(\ref{NI})), which has been shown to
exhibit the minimal phase diffusion \cite{martin2012quantum}. However, $\gamma$ cannot be increased indefinitely, as the geometry of the ring trap constrains its maximum permissible value. Specifically, the initial droplet size increases with $\gamma$, and to ensure that the split daughter droplets remain distinguishable and separated, we restrict ourselves to values of $\gamma$ for which the droplet size is smaller than half the ring's circumference. Consequently, increasing the radius helps to enhance the sensitivity in two ways: (1) by increasing the accumulated Sagnac phase, and (2) by allowing access to larger values of $\gamma $, which further improves the sensitivity. That said, a larger ring also extends the interferometric cycle time, potentially making the system more susceptible to noise and atom loss.

Apart from $\gamma $, $N$ also plays a crucial role in
determining the contrast and sensitivity. In the LHY-superfluid regime, the
increase of $N$ leads to a reduction in both the contrast and sensitivity. However, in the MF-dominated regime, the increase of $N$ enhances the contrast and maximum sensitivity, highlighting the importance of optimising the atom number for the improved interferometric performance.

\section{Conclusion}

\label{SecIV} We have investigated the behavior of QDs in
the harmonic-oscillator-trapped and ring-shaped setups, demonstrating
their suitability for the realisation of the QD-based interferometric
schemes. The analysis highlights the significance of the atom number ($N$)
and relative mean-field interaction strength ($\gamma $) in determining QD
properties in both the HO-trapped and ring-shaped setups. Specifically, $N$
and $\gamma $ affect the width of the QD in the ring configuration, and the
crossover between QDs and HO-determined bound states. Building on this foundation, we\ have developed the schemes for the HO-confined ring-shaped QD-based interferometers. For a given barrier width, we have identified the barrier height necessary for the $50:50$ splitting, by analysing the transmission coefficient for the QD collision with the first barrier. The relative atom number observed after the recombination of the daughter droplets provides an appropriate measure for the
interferometer's phase sensitivity. Additionally, we examined the sensitivity of both interferometers to variations in the barrier position.

A key outcome of our study is the demonstration of the HO-trapped QD interferometer as a tilt-meter and target detector, and the ring-shaped QD interferometer as a Sagnac interferometer. The tilt-meter achieves the maximum sensitivity at high values of $\gamma $, though this result comes with poor contrast. A better contrast occurs at lower values of $\gamma $ -- specifically, in the region of the crossover between the QDs and
HO-determined bound states. On the other hand, for target detection, the sensitivity peaks at $\gamma=25$, with a desirable contrast of $90\%$. The maximum contrast also occurs near the crossover region, similar to the tilt-meter case. Thus, optimising the performance of the tilt-meter or target detection requires maintaining a careful trade-off between the contrast and sensitivity. In the Sagnac interferometer, both the contrast and sensitivity increase with the growth of $\gamma $, although the largest attainable value of $\gamma $ is constrained by the ring's size. At sufficiently large values of $\gamma $, the QD width approaches half the ring circumference, leading to reduced distinguishability between the split daughter droplets. A larger ring helps to improve the sensitivity by increasing the accumulated Sagnac phase and also by enabling access to higher values of $\gamma $. However, the use of larger rings leads to a longer interferometric cycle time, which may increase the susceptibility to noise and atom loss. An promising direction for future research may be the study of the number statistics of the QD interferometer, using methods such as the truncated-Wigner approximation \cite{blakie2008dynamics,martin2012quantum,haine2018quantum}.
Because large-atom-number QDs can be created with the minimal density, which provides weak interactions in them, compared to bright solitons, hence one may expect less degradation of the sensitivity due to atom-number fluctuations.

Regarding the experimental feasibility of the setups developed here, pure 1D
QDs have not yet been observed experimentally, while robust 3D QDs have been
created in Bose-Bose mixtures and dipolar BECs, both in the free space and
in elongated (cigar-shaped) potentials \cite%
{cabrera2018quantum,chomaz2019long,cheiney2018bright}. Achieving a true 1D
QD requires strong transverse confinement, such that ratio $\xi $ of the MF interaction energy to the transverse confinement energy remains
$<0.03$ \cite{zin2018quantum}. In our analysis, we ensured this condition by
imposing the tight transverse confinement. Specifically, we consider a
mixture of two hyperfine states in ${}^{39}\mathrm{K}$ atoms, confined by the
transverse HO trap with frequency $\omega _{\perp }=2\pi \times 500$ Hz. In
this setting, the intraspecies scattering lengths are $a_{11}=a_{22}=300a_{0}
$, while the interspecies one may be varied between $-a_{11}$ and $-0.6a_{11}
$, thus tuning parameter $\gamma $ from $0$ to $\approx 1000.$ In this range
of the values of $a_{12}$, the respective LHY amended GPE provides a reliable QD model \cite{parisi2019liquid}.

\section{References}

\bibliographystyle{iopart-num}
\bibliography{ref}

@article{petrov2015quantum,
	title={Quantum mechanical stabilization of a collapsing Bose-Bose mixture},
	author={Petrov, DS},
	journal={Physical review letters},
	volume={115},
	number={15},
	pages={155302},
	year={2015},
	publisher={APS}
}

@article{jorgensen2018dilute,
	title={Dilute fluid governed by quantum fluctuations},
	author={J{\o}rgensen, Nils B and Bruun, Georg M and Arlt, Jan J},
	journal={Physical review letters},
	volume={121},
	number={17},
	pages={173403},
	year={2018},
	publisher={APS}
}

@article{skov2021observation,
	title={Observation of a lee-huang-yang fluid},
	author={Skov, Thomas G and Skou, Magnus G and J{\o}rgensen, Nils B and Arlt, Jan J},
	journal={Physical Review Letters},
	volume={126},
	number={23},
	pages={230404},
	year={2021},
	publisher={APS}
}

@article{luo2021new,
	title={A new form of liquid matter: Quantum droplets},
	author={Luo, Zhi-Huan and Pang, Wei and Liu, Bin and Li, Yong-Yao and Malomed, Boris A},
	journal={Frontiers of Physics},
	volume={16},
	pages={1--21},
	year={2021},
	publisher={Springer}
}

@book{pitaevskii2016bose,
	title={Bose-Einstein condensation and superfluidity},
	author={Pitaevskii, Lev and Stringari, Sandro},
	volume={164},
	year={2016},
	publisher={Oxford University Press}
}

@book{pethick2008bose,
	title={Bose--Einstein condensation in dilute gases},
	author={Pethick, Christopher J and Smith, Henrik},
	year={2008},
	publisher={Cambridge university press}
}

@article{petrov2016ultradilute,
	title={Ultradilute low-dimensional liquids},
	author={Petrov, DS and Astrakharchik, GE},
	journal={Physical review letters},
	volume={117},
	number={10},
	pages={100401},
	year={2016},
	publisher={APS}
}

@article{zin2018quantum,
	title={Quantum Bose-Bose droplets at a dimensional crossover},
	author={Zin, Pawe{\l} and Pylak, Maciej and Wasak, Tomasz and Gajda, Mariusz and Idziaszek, Zbigniew},
	journal={Physical Review A},
	volume={98},
	number={5},
	pages={051603},
	year={2018},
	publisher={APS}
}

@article{lee1957eigenvalues,
	title={Eigenvalues and eigenfunctions of a Bose system of hard spheres and its low-temperature properties},
	author={Lee, Tsin D and Huang, Kerson and Yang, Chen N},
	journal={Physical Review},
	volume={106},
	number={6},
	pages={1135},
	year={1957},
	publisher={APS}
}

@article{kadau2016observing,
	title={Observing the Rosensweig instability of a quantum ferrofluid},
	author={Kadau, Holger and Schmitt, Matthias and Wenzel, Matthias and Wink, Clarissa and Maier, Thomas and Ferrier-Barbut, Igor and Pfau, Tilman},
	journal={Nature},
	volume={530},
	number={7589},
	pages={194--197},
	year={2016},
	publisher={Nature Publishing Group UK London}
}

@article{ferrier2016observation,
	title={Observation of quantum droplets in a strongly dipolar Bose gas},
	author={Ferrier-Barbut, Igor and Kadau, Holger and Schmitt, Matthias and Wenzel, Matthias and Pfau, Tilman},
	journal={Physical review letters},
	volume={116},
	number={21},
	pages={215301},
	year={2016},
	publisher={APS}
}

@article{chomaz2019long,
	title={Long-lived and transient supersolid behaviors in dipolar quantum gases},
	author={Chomaz, L and Petter, D and Ilzh{\"o}fer, P and Natale, G and Trautmann, A and Politi, C and Durastante, G and Van Bijnen, RMW and Patscheider, A and Sohmen, M and others},
	journal={Physical Review X},
	volume={9},
	number={2},
	pages={021012},
	year={2019},
	publisher={APS}
}

@article{chomaz2016quantum,
	title={Quantum-fluctuation-driven crossover from a dilute Bose-Einstein condensate to a macrodroplet in a dipolar quantum fluid},
	author={Chomaz, L and Baier, S and Petter, D and Mark, MJ and W{\"a}chtler, F and Santos, Luis and Ferlaino, F},
	journal={Physical Review X},
	volume={6},
	number={4},
	pages={041039},
	year={2016},
	publisher={APS}
}

@article{cabrera2018quantum,
	title={Quantum liquid droplets in a mixture of Bose-Einstein condensates},
	author={Cabrera, CR and Tanzi, L and Sanz, J and Naylor, B and Thomas, P and Cheiney, P and Tarruell, Leticia},
	journal={Science},
	volume={359},
	number={6373},
	pages={301--304},
	year={2018},
	publisher={American Association for the Advancement of Science}
}

@article{semeghini2018self,
	title={Self-bound quantum droplets of atomic mixtures in free space},
	author={Semeghini, G and Ferioli, G and Masi, Leonardo and Mazzinghi, C and Wolswijk, Louise and Minardi, F and Modugno, M and Modugno, G and Inguscio, M and Fattori, M},
	journal={Physical review letters},
	volume={120},
	number={23},
	pages={235301},
	year={2018},
	publisher={APS}
}

@article{cheiney2018bright,
	title={Bright soliton to quantum droplet transition in a mixture of Bose-Einstein condensates},
	author={Cheiney, P and Cabrera, CR and Sanz, J and Naylor, B and Tanzi, L and Tarruell, L},
	journal={Physical review letters},
	volume={120},
	number={13},
	pages={135301},
	year={2018},
	publisher={APS}
}

@article{astrakharchik2018dynamics,
	title={Dynamics of one-dimensional quantum droplets},
	author={Astrakharchik, GE and Malomed, Boris A},
	journal={Physical Review A},
	volume={98},
	number={1},
	pages={013631},
	year={2018},
	publisher={APS}
}

@article{parisi2019liquid,
	title={Liquid state of one-dimensional Bose mixtures: A quantum Monte Carlo study},
	author={Parisi, Luca and Astrakharchik, GE and Giorgini, Stefano},
	journal={Physical review letters},
	volume={122},
	number={10},
	pages={105302},
	year={2019},
	publisher={APS}
}

@article{parisi2020quantum,
	title={Quantum droplets in one-dimensional Bose mixtures: A quantum Monte Carlo study},
	author={Parisi, Luca and Giorgini, Stephano},
	journal={Physical Review A},
	volume={102},
	number={2},
	pages={023318},
	year={2020},
	publisher={APS}
}

@article{mithun2020modulational,
	title={Modulational instability, inter-component asymmetry, and formation of quantum droplets in one-dimensional binary Bose gases},
	author={Mithun, Thudiyangal and Maluckov, Aleksandra and Kasamatsu, Kenichi and Malomed, Boris A and Khare, Avinash},
	journal={Symmetry},
	volume={12},
	number={1},
	pages={174},
	year={2020},
	publisher={MDPI}
}

@article{debnath2021investigation,
	title={Investigation of quantum droplets: An analytical approach},
	author={Debnath, Argha and Khan, Ayan},
	journal={Annalen der Physik},
	volume={533},
	number={3},
	pages={2000549},
	year={2021},
	publisher={Wiley Online Library}
}

@article{otajonov2019stationary,
	title={Stationary and dynamical properties of one-dimensional quantum droplets},
	author={Otajonov, Sherzod R and Tsoy, Eduard N and Abdullaev, Fatkhulla Kh},
	journal={Physics Letters A},
	volume={383},
	number={34},
	pages={125980},
	year={2019},
	publisher={Elsevier}
}

@article{tylutki2020collective,
	title={Collective excitations of a one-dimensional quantum droplet},
	author={Tylutki, Marek and Astrakharchik, Grigori E and Malomed, Boris A and Petrov, Dmitry S},
	journal={Physical Review A},
	volume={101},
	number={5},
	pages={051601},
	year={2020},
	publisher={APS}
}

@article{englezos2023correlated,
	title={Correlated dynamics of collective droplet excitations in a one-dimensional harmonic trap},
	author={Englezos, IA and Mistakidis, Simeon I and Schmelcher, P},
	journal={Physical Review A},
	volume={107},
	number={2},
	pages={023320},
	year={2023},
	publisher={APS}
}

@article{du2023ground,
	title={Ground-state properties and Bogoliubov modes of a harmonically trapped one-dimensional quantum droplet},
	author={Du, Xucong and Fei, Yifan and Chen, Xiao-Long and Zhang, Yunbo},
	journal={Physical Review A},
	volume={108},
	number={3},
	pages={033312},
	year={2023},
	publisher={APS}
}

@article{zezyulin2023quasi,
	title={Quasi-one-dimensional harmonically trapped quantum droplets},
	author={Zezyulin, Dmitry A},
	journal={Physical Review A},
	volume={107},
	number={4},
	pages={043307},
	year={2023},
	publisher={APS}
}

@article{pathak2022dynamics,
	title={Dynamics of quantum droplets in an external harmonic confinement},
	author={Pathak, Maitri R and Nath, Ajay},
	journal={Scientific Reports},
	volume={12},
	number={1},
	pages={6904},
	year={2022},
	publisher={Nature Publishing Group UK London}
}

@article{zhao2021discrete,
	title={Discrete quantum droplets in one-dimensional optical lattices},
	author={Zhao, Fei-yan and Yan, Zi-teng and Cai, Xiao-yan and Li, Chao-long and Chen, Gui-lian and He, He-xiang and Liu, Bin and Li, Yong-yao},
	journal={Chaos, Solitons \& Fractals},
	volume={152},
	pages={111313},
	year={2021},
	publisher={Elsevier}
}

@article{zhou2019dynamics,
	title={Dynamics of quantum droplets in a one-dimensional optical lattice},
	author={Zhou, Zheng and Yu, Xi and Zou, Yu and Zhong, Honghua},
	journal={Communications in Nonlinear Science and Numerical Simulation},
	volume={78},
	pages={104881},
	year={2019},
	publisher={Elsevier}
}

@article{nie2023spectra,
	title={Spectra and dynamics of quantum droplets in an optical lattice},
	author={Nie, Yuhang and Zheng, Jun-Hui and Yang, Tao},
	journal={Physical Review A},
	volume={108},
	number={5},
	pages={053310},
	year={2023},
	publisher={APS}
}

@article{flynn2024harmonically,
	title={Harmonically trapped imbalanced quantum droplets},
	author={Flynn, TA and Keepfer, NA and Parker, NG and Billam, TP},
	journal={Physical Review Research},
	volume={6},
	number={1},
	pages={013209},
	year={2024},
	publisher={APS}
}

@article{hu2023scattering,
	title={Scattering of one-dimensional quantum droplets by a reflectionless potential well},
	author={Hu, Xiaoxiao and Li, Zhiqiang and Guo, Yu and Chen, Yajiang and Luo, Xiaobing},
	journal={Physical Review A},
	volume={108},
	number={5},
	pages={053306},
	year={2023},
	publisher={APS}
}

@article{debnath2023interaction,
	title={Interaction of one-dimensional quantum droplets with potential wells and barriers},
	author={Debnath, Argha and Khan, Ayan and Malomed, Boris},
	journal={Communications in Nonlinear Science and Numerical Simulation},
	volume={126},
	pages={107457},
	year={2023},
	publisher={Elsevier}
}

@article{chiquillo2019low,
	title={Low-dimensional self-bound quantum Rabi-coupled bosonic droplets},
	author={Chiquillo, Emerson},
	journal={Physical Review A},
	volume={99},
	number={5},
	pages={051601},
	year={2019},
	publisher={APS}
}

@article{xiong2021self,
	title={Self-bound quantum droplet with internal stripe structure in one-dimensional spin-orbit-coupled Bose gas},
	author={Xiong, Yuncheng and Yin, Lan},
	journal={Chinese Physics Letters},
	volume={38},
	number={7},
	pages={070301},
	year={2021},
	publisher={IOP Publishing}
}

@article{singh2020modulational,
	title={Modulational instability in a one-dimensional spin--orbit coupled Bose--Bose mixture},
	author={Singh, Dheerendra and Parit, Mithilesh K and Raju, Thokala Soloman and Panigrahi, Prasanta K},
	journal={Journal of Physics B: Atomic, Molecular and Optical Physics},
	volume={53},
	number={24},
	pages={245001},
	year={2020},
	publisher={IOP Publishing}
}

@article{chandramouli2024dispersive,
	title={Dispersive shock waves in a one-dimensional droplet-bearing environment},
	author={Chandramouli, Sathyanarayanan and Mistakidis, Simeon I and Katsimiga, Garyfallia C and Kevrekidis, Panayotis G},
	journal={Physical Review A},
	volume={110},
	number={2},
	pages={023304},
	year={2024},
	publisher={APS}
}

@article{kevrekidis2004avoiding,
	title={Avoiding infrared catastrophes in trapped Bose-Einstein condensates},
	author={Kevrekidis, PG and Theocharis, G and Frantzeskakis, DJ and Trombettoni, A},
	journal={Physical Review A—Atomic, Molecular, and Optical Physics},
	volume={70},
	number={2},
	pages={023602},
	year={2004},
	publisher={APS}
}

@article{helm2015sagnac,
	title={Sagnac interferometry using bright matter-wave solitons},
	author={Helm, JL and Cornish, SL and Gardiner, SA},
	journal={Physical review letters},
	volume={114},
	number={13},
	pages={134101},
	year={2015},
	publisher={APS}
}

@article{marchant2013controlled,
	title={Controlled formation and reflection of a bright solitary matter-wave},
	author={Marchant, AL and Billam, TP and Wiles, TP and Yu, MMH and Gardiner, SA and Cornish, SL},
	journal={Nature communications},
	volume={4},
	number={1},
	pages={1865},
	year={2013},
	publisher={Nature Publishing Group UK London}
}

@article{helm2012bright,
	title={Bright matter-wave soliton collisions at narrow barriers},
	author={Helm, John L and Billam, Thomas P and Gardiner, Simon A},
	journal={Physical Review A—Atomic, Molecular, and Optical Physics},
	volume={85},
	number={5},
	pages={053621},
	year={2012},
	publisher={APS}
}

@article{polo2013soliton,
	title={Soliton-based matter-wave interferometer},
	author={Polo, Juan and Ahufinger, Ver{\`o}nica},
	journal={Physical Review A—Atomic, Molecular, and Optical Physics},
	volume={88},
	number={5},
	pages={053628},
	year={2013},
	publisher={APS}
}

@article{martin2012quantum,
	title={Quantum dynamics of atomic bright solitons under splitting and recollision, and implications for interferometry},
	author={Martin, AD and Ruostekoski, J},
	journal={New Journal of physics},
	volume={14},
	number={4},
	pages={043040},
	year={2012},
	publisher={IOP Publishing}
}

@article{helm2014splitting,
	title={Splitting bright matter-wave solitons on narrow potential barriers: Quantum to classical transition and applications to interferometry},
	author={Helm, JL and Rooney, SJ and Weiss, Christoph and Gardiner, SA},
	journal={Physical Review A},
	volume={89},
	number={3},
	pages={033610},
	year={2014},
	publisher={APS}
}

@article{wales2020splitting,
	title={Splitting and recombination of bright-solitary-matter waves},
	author={Wales, Oliver J and Rakonjac, Ana and Billam, Thomas P and Helm, John L and Gardiner, Simon A and Cornish, Simon L},
	journal={Communications Physics},
	volume={3},
	number={1},
	pages={51},
	year={2020},
	publisher={Nature Publishing Group UK London}
}

@article{sakaguchi2016matter,
	title={Matter-wave soliton interferometer based on a nonlinear splitter},
	author={Sakaguchi, Hidetsugu and Malomed, Boris A},
	journal={New Journal of Physics},
	volume={18},
	number={2},
	pages={025020},
	year={2016},
	publisher={IOP Publishing}
}

@article{mcdonald2014bright,
	title={Bright solitonic matter-wave interferometer},
	author={McDonald, Gordon D and Kuhn, Carlos CN and Hardman, Kyle S and Bennetts, Shayne and Everitt, Patrick J and Altin, Paul A and Debs, John E and Close, John D and Robins, Nicholas P},
	journal={Physical review letters},
	volume={113},
	number={1},
	pages={013002},
	year={2014},
	publisher={APS}
}

@article{cuevas2013interactions,
	title={Interactions of solitons with a Gaussian barrier: splitting and recombination in quasi-one-dimensional and three-dimensional settings},
	author={Cuevas, J and Kevrekidis, PG and Malomed, BA and Dyke, P and Hulet, RG},
	journal={New Journal of Physics},
	volume={15},
	number={6},
	pages={063006},
	year={2013},
	publisher={IOP Publishing}
}

@article{sun2014mean,
	title={Mean-field analog of the Hong-Ou-Mandel experiment with bright solitons},
	author={Sun, Zhi-Yuan and Kevrekidis, Panayotis G and Kr{\"u}ger, Peter},
	journal={Physical Review A},
	volume={90},
	number={6},
	pages={063612},
	year={2014},
	publisher={APS}
}

@article{haine2018quantum,
	title={Quantum noise in bright soliton matterwave interferometry},
	author={Haine, Simon A},
	journal={New Journal of Physics},
	volume={20},
	number={3},
	pages={033009},
	year={2018},
	publisher={IOP Publishing}
}

@article{lehtovaara2007solution,
	title={Solution of time-independent Schr{\"o}dinger equation by the imaginary time propagation method},
	author={Lehtovaara, Lauri and Toivanen, Jari and Eloranta, Jussi},
	journal={Journal of Computational Physics},
	volume={221},
	number={1},
	pages={148--157},
	year={2007},
	publisher={Elsevier}
}

@article{ernst2010probing,
	title={Probing superfluids in optical lattices by momentum-resolved Bragg spectroscopy},
	author={Ernst, Philipp T and G{\"o}tze, S{\"o}ren and Krauser, Jasper S and Pyka, Karsten and L{\"u}hmann, Dirk-S{\"o}ren and Pfannkuche, Daniela and Sengstock, Klaus},
	journal={Nature Physics},
	volume={6},
	number={1},
	pages={56--61},
	year={2010},
	publisher={Nature Publishing Group UK London}
}

@article{kozuma1999coherent,
	title={Coherent splitting of Bose-Einstein condensed atoms with optically induced Bragg diffraction},
	author={Kozuma, M and Deng, Lu and Hagley, Edward W and Wen, J and Lutwak, R and Helmerson, Kristian and Rolston, SL and Phillips, William D},
	journal={Physical Review Letters},
	volume={82},
	number={5},
	pages={871},
	year={1999},
	publisher={APS}
}

@article{burke2009scalable,
	title={Scalable Bose-Einstein-condensate Sagnac interferometer in a linear trap},
	author={Burke, JHT and Sackett, CA},
	journal={Physical Review A—Atomic, Molecular, and Optical Physics},
	volume={80},
	number={6},
	pages={061603},
	year={2009},
	publisher={APS}
}

@article{gautier2022accurate,
	title={Accurate measurement of the Sagnac effect for matter waves},
	author={Gautier, Romain and Guessoum, Mohamed and Sidorenkov, Leonid A and Bouton, Quentin and Landragin, Arnaud and Geiger, Remi},
	journal={Science Advances},
	volume={8},
	number={23},
	pages={eabn8009},
	year={2022},
	publisher={American Association for the Advancement of Science}
}

@article{bera2022quantum,
	title={Quantum sensing with sub-Planck structures for the dynamics of Bose-Einstein condensate in presence of engineered potential barriers inside a harmonic trap},
	author={Bera, Jayanta and Halder, Barun and Ghosh, Suranjana and Lee, Ray-Kuang and Roy, Utpal},
	journal={Physics Letters A},
	volume={453},
	pages={128484},
	year={2022},
	publisher={Elsevier}
}

@article{raghav2024dispersion,
	title={Dispersion-managed elliptical atomtronics for interferometry},
	author={Raghav, Sriganapathy and Ghosh, Suranjana and Bera, Jayanta and Roy, Utpal},
	journal={The European Physical Journal Plus},
	volume={139},
	number={11},
	pages={1--11},
	year={2024},
	publisher={Springer}
}

@article{raghav2025nonlinearity,
	title={Nonlinearity mediated miscibility dynamics of mass-imbalanced binary Bose--Einstein condensate for circular atomtronics},
	author={Raghav, Sriganapathy and Ghosh, Suranjana and Halder, Barun and Roy, Utpal},
	journal={Physica D: Nonlinear Phenomena},
	pages={134558},
	year={2025},
	publisher={Elsevier}
}

@article{bera2020matter,
	title={Matter-wave fractional revivals in a ring waveguide},
	author={Bera, Jayanta and Ghosh, Suranjana and Salasnich, Luca and Roy, Utpal},
	journal={Physical Review A},
	volume={102},
	number={6},
	pages={063323},
	year={2020},
	publisher={APS}
}

@book{MalomedMS,
	author = {Malomed, Boris A.},
	title = {Multidimensional Solitons},
	publisher = {AIP Publishing LLC},
	isbn = {978-0-7354-2509-5},
	doi = {10.1063/9780735425118},
}

@article{d2019observation,
	title={Observation of quantum droplets in a heteronuclear bosonic mixture},
	author={D'Errico, C and Burchianti, A and Prevedelli, M and Salasnich, L and Ancilotto, F and Modugno, M and Minardi, F and Fort, C},
	journal={Physical Review Research},
	volume={1},
	number={3},
	pages={033155},
	year={2019},
	publisher={APS}
}

@article{blakie2008dynamics,
	title={Dynamics and statistical mechanics of ultra-cold Bose gases using c-field techniques},
	author={Blakie, P Blair and Bradley, AS and Davis, MJ and Ballagh, RJ and Gardiner, CW},
	journal={Advances in Physics},
	volume={57},
	number={5},
	pages={363--455},
	year={2008},
	publisher={Taylor \& Francis}
}

@article{berrada2013integrated,
  title={Integrated mach--zehnder interferometer for bose--einstein condensates},
  author={Berrada, Tarik and Van Frank, Sandrine and B{\"u}cker, Robert and Schumm, Thorsten and Schaff, J-F and Schmiedmayer, J{\"o}rg},
  journal={Nature communications},
  volume={4},
  number={1},
  pages={2077},
  year={2013},
  publisher={Nature Publishing Group UK London}
}

@article{javanainen1997phase,
  title={Phase and phase diffusion of a split Bose-Einstein condensate},
  author={Javanainen, Juha and Wilkens, Martin},
  journal={Physical Review Letters},
  volume={78},
  number={25},
  pages={4675},
  year={1997},
  publisher={APS}
}

@article{carr2002dynamics,
  title={Dynamics of a matter-wave bright soliton in an expulsive potential},
  author={Carr, Lincoln D and Castin, Yvan},
  journal={Physical Review A},
  volume={66},
  number={6},
  pages={063602},
  year={2002},
  publisher={APS}
}

@article{salasnich2017bright,
  title={Bright solitons in ultracold atoms},
  author={Salasnich, Luca},
  journal={Optical and Quantum Electronics},
  volume={49},
  number={12},
  pages={409},
  year={2017},
  publisher={Springer}
}

@article{damgaard2021scattering,
  title={Scattering of matter wave solitons on localized potentials},
  author={Damgaard Hansen, Sidse and Nygaard, Nicolai and M{\o}lmer, Klaus},
  journal={Applied Sciences},
  volume={11},
  number={5},
  pages={2294},
  year={2021},
  publisher={MDPI}
}

@article{konotop1996interaction,
  title={Interaction of a soliton with point impurities in an inhomogeneous, discrete nonlinear Schr{\"o}dinger system},
  author={Konotop, VV and Cai, David and Salerno, Mario and Bishop, AR and Gr{\o}nbech-Jensen, Niels},
  journal={Physical Review E},
  volume={53},
  number={6},
  pages={6476},
  year={1996},
  publisher={APS}
}

\end{document}